\documentclass{article}
\usepackage{latexsym}
\usepackage{graphicx}    
\begin{document}
\title{Exact diagonalization of the S = 1/2 XY ferromagnet on
finite bcc lattices to estimate T = 0 properties on the infinite
lattice}
\author{J. Schulenburg $^1$, 
 J. S. Flynn $^2$, 
 D. D. Betts $^2$ 
 and J. Richter $^1$ 
\\
\and $^1$ Institute f\"ur Theoretische Physik, Univesit\"at
Magdeburg, Germany \and $^2$ Department
of Physics, Dalhousie University, Halifax, N.S., Canada
}
%
\date{Dec 2000}
%
\maketitle
\begin{abstract}
 In this paper finite bcc lattices are defined by a
triple of vectors in two different ways - upper triangular lattice
form and compact form. In Appendix A are lists of some 260
distinct and useful bcc lattices of 9 to 32 vertices. The energy
and magnetization of the S = 1/2 XY ferromagnet have been computed
on these bcc lattices in the lowest states for $\mathrm{S_z}$ = 0,
1/2, 1 and 3/2. These data are studied statistically to fit the
first three terms of the appropriate finite lattice scaling
equations. Our estimates of the T = 0 energy and magnetization
agree very well with spin wave and series expansion estimates.
\end{abstract} 
%
%
\section{Introduction}
\label{intro} The physics of quantum spin systems on lattices of d
= 1, 2 and 3 dimensions has been much studied for several decades.
Several different models have been studied including the
Heisenberg antiferromagnet, the t-J model, the Hubbard model, the
spin fermion model, the spin-orbital model and the XY ferromagnet.
There are several different methods to study quantum spin models,
especially in one dimension. It is easiest to study
three-dimensional spin models precisely at high temperatures, and
zero temperature is second easiest by some methods. While
experiments never reach T = 0, several do work at extremely low
temperature. In three dimensions, as opposed to two dimensions,
the zero temperature properties are very similar to those at very
low temperature.

Useful methods to study quantum spin models such as the S = 1/2
Heisenberg and XY models at T = 0 in three dimensions (d = 3)
include series expansion, \cite{Singh89}, \cite{Oitmaa94}, spin
wave, \cite{Anderson52}, \cite{Kubo52}, \cite{Gomez-Santos87}, and
variational \cite{Suzuki78} methods (many other references could
be cited). The S = 1/2 Heisenberg antiferromagnet is generally
regarded as the most important quantum spin model on a lattice,
and studies of this model in one, two and three dimensions are
very numerous. However, the spin one-half XY model is also quite
important as it is the simplest fully quantum mechanical lattice
model. Recent (1999) examples have used quantum Monte Carlo and/or
finite lattice method on the square lattice \cite{Hamer99},
\cite{Sandvik99}, \cite{Betts99}. In this d = 3, T = 0 regime we
use the latest method - exact diagonalization of quantum spin
models on finite three dimensional lattices \cite{BettsStewart97},
\cite{BettsStewartFlynn97}, \cite{Betts98}.

Each finite lattice in three dimensions can be derived from one or
more parallelepipeds. Such a parallelepiped can be defined by
three edge vectors in such a length and direction as to have a
lattice vertex on each of the eight parallelepiped corners. Thus a
set of identical, regularly packed parallelepipeds will completely
fill the infinite lattice - a "three dimensional tiling". A d = 3
lattice is formed by identifying each pair of opposite faces of
the parallelepiped as being one and the same face. More detail is
described in Section 2 below. In particular, the very useful {\em
upper triangular lattice form} is described in Section 2.

Section 3 describes the generation of finite bcc lattices that can
be used for {\em ferromagnetic} models. The number of such bcc
lattices of N $\le 32$ vertices is an order of magnitude greater
than the number of {\em bipartite} lattices needed for the study
of an antiferromagnetic model. This section also explains the {\em
geometric imperfection}, $\mathrm{I_G}$, of finite lattices. If
$\mathrm{I_G}$/N for a specific finite lattice is too large, that
finite lattice is not used. An N = 15 example of a bcc lattice is
shown in Fig. 1. The description of all the useful bcc lattices
used is listed in Appendix A. Each distinct lattice is labelled
N.i.

Section 4 describes the computation of physical properties of the
S = 1/2 XY ferromagnet on finite bcc lattices in the ground state
and in the first excited state. Thousands of hours on the powerful
computer at the University of Magdeburg were used to compute the
physical properties of the S = 1/2 XY ferromagnet on all the
finite bcc lattices used. Specifically we needed about seventy
hours to calculate the eigenstate of the largest Hamiltonian (N =
32) using an SGI Power-Challenge with 2GB of memory. The detailed
results are shown in Appendix B.

In Section 5 statistical analysis estimates the T = 0 properties
of the S = 1/2 XY ferromagnet on the infinite body-centred cubic
lattice. The data for physical properties of this model on finite
bcc lattices are statistically analysed in order to obtain the
estimates of the first two or three coefficients of the
appropriate finite lattice scaling equation. The results are
displayed in Tables 2, 3, 4 and 5.

Summary, conclusions and outlook are described in Section 6.

\section{Definition of finite bcc lattices}
\label{defn} This process has been described in an earlier paper
\cite{Betts98} but some aspects, new or repeated, will be
displayed here. The infinite bcc lattice can be defined by any
three of the four primitive vectors:
\begin{eqnarray}
\label{primvects}
\nonumber
\mathbf{a}_1 = (1,1,1), \hspace{1em} \mathbf{a}_2 = (1,1,-1), \\
\mathbf{a}_3 = (1,-1,1) \hspace{1em} \mathrm{and} \hspace{1em}
\mathbf{a}_4 = (-1,1,1)
\end{eqnarray}
The infinite bcc lattice can be "filled" (the three dimensional
analogy of "tiled" in two dimensions) by any one of several sets
of an ascending number of identical parallelepipeds. Such a
parallelepiped is defined by three edge vectors,
\begin{equation}
\label{edgevects}
\mathbf{L}_\alpha = \sum_{\beta=1}^{3} n_{\alpha\beta} \mathbf{a}_{\beta},
\end{equation}
in which each $n_{\alpha\beta}$ is an integer. Each of three
integers $n_{\alpha 1}$, $n_{\alpha 2}$ and $n_{\alpha 3}$ are odd
or each are even. A finite bcc lattice can be derived from any bcc
parallelepiped defined by (\ref{edgevects}) by applying a full set
of three periodic boundary conditions. That is, each pair of the
opposite faces of the parallelepiped are defined to be identical.
In other words, each finite lattice is the three dimensional
analogue of a two dimensional torus.

\begin{figure}
\begin{center}
\resizebox{0.66\textwidth}{!}{
  \includegraphics{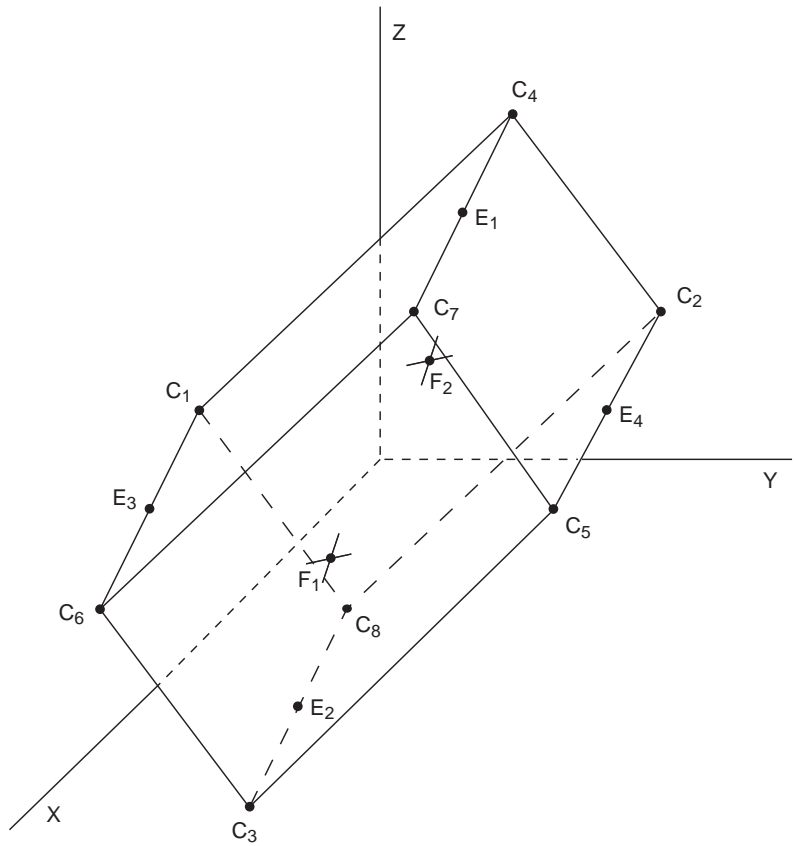}
 }
\end{center}
\caption{A parallelepiped of N = 15 vertices embedded in an
infinite bcc lattice. By identifying each pair of opposite faces
this parallelepiped becomes a finite bcc lattice of fifteen
vertices.}
\label{N15lattice}       
\end{figure}

Figure 1 provides a concrete example, a parallelepiped that is
turned into a finite bcc lattice with fifteen distinct vertices.
The original parallelepiped's eight corners are labeled
$\mathrm{C}_1$, $\mathrm{C}_2$ ... $\mathrm{C}_8$. The four
vertices on four of the twelve edges of the parallelepiped are
labeled $\mathrm{E}_1$, $\mathrm{E}_2$, $\mathrm{E}_3$ and
$\mathrm{E}_4$, and the two vertices on one pair of opposite faces
of the parallelepiped are labeled $\mathrm{F}_1$ and
$\mathrm{F}_2$. This parallelepiped contains also twelve {\em
internal} vertices we could label as A, B, D, G, H, I, J, K, L, M,
N and P. However, it appears simpler not to show these internal
vertices in Figure 1. The three defining edge vectors from
$\mathrm{C}_8$ to $\mathrm{C}_1$, $\mathrm{C}_2$ and
$\mathrm{C}_3$ respectively are $\mathbf{L}_1 = (1,-1,3)$,
$\mathbf{L}_2 = (-1,3,3)$ and $\mathbf{L}_3 = (4,2,0)$. The volume
of this parallelepiped, $\mathrm{V} = \mathbf{L}_3 \cdot
(\mathbf{L}_2 \times \mathbf{L}_1) = 60$.

By identifying the three pairs of opposite faces of this
parallelepiped, i.e. applying periodic boundary conditions, the
two face vertices $\mathrm{F}_1$ and $\mathrm{F}_2$ become F, the
four edge vertices, $\mathrm{E}_1$, $\mathrm{E}_2$, $\mathrm{E}_3$
and $\mathrm{E}_4$ become one other vertex E, and the eight corner
vertices $\mathrm{C}_1$, $\mathrm{C}_2$, ... $\mathrm{C}_8$ are
all now the single vertex C. The finite bcc lattice thus formed
contains fifteen vertices A, B, C, D, E, ... P. A finite lattice
has no faces, edges nor corners, and all vertices have the same
geometric environment.

In fact, a typical finite lattice can be derived from any one of
several parallelepipeds and thus it might appear to be several
distinct finite lattices. However, it has been proved by Lyness et
al \cite{Lyness91} that, in effect, each parallelepiped in any
number of dimensions, d, defined by a matrix in upper triangular
lattice form (utlf) will, upon complete application of periodic
boundary conditions, form a unique finite lattice.

A d-dimensional utlf matrix, $\mathrm{L}^t$, must satisfy the
following criteria:
\begin{eqnarray}
\nonumber \mathrm{L}^t_{ii} \ge 1  & \hspace{2em} &  i =
1,2,\ldots,d
\\ \mathrm{L}^t_{ij} = 0  & \hspace{2em} & 1 \le j < i \le d \\ \nonumber
 \mathrm{L}^t_{ij} \in [ 0,\mathrm{L}^t_{jj} ) & \hspace{2em} &  1 \le i < j \le d \\
\nonumber \mathrm{L}^t_{i+1,i+1} \ge \mathrm{L}^t_{ii} &
\hspace{2em} &
\end{eqnarray}

In particular, in an unbounded three dimensional lattice, the
defining utlf matrix is
\begin{equation}
\label{utlfmatrix}
 \mathrm{L}^t = \left(
\begin{array}{c}
\mathbf{L}^t_1 \\ \mathbf{L}^t_2 \\ \mathbf{L}^t_3 \\
\end{array}
\right)
\end{equation}
where the $\mathbf{L}^t_i$ are the defining vectors.

Furthermore, on the infinite bcc lattice defined by the above
primitive defining vectors $\mathbf{a}_{\beta}$, the number of
vertices, N, in a finite utlf description of a finite bcc lattice
requires
\begin{equation}
\label{nfromls}
\mathrm{L}^t_{11}\mathrm{L}^t_{22}\mathrm{L}^t_{33} = 4
\mathrm{N}.
\end{equation}
Since $\mathbf{L}^t_3$ lies on the Z axis and $\mathbf{L}^t_2$
lies in the YZ plane all of their components are even integers. In
particular, the utlf vectors defining the N = 15 finite bcc
lattice 15.28, described above, are $\mathbf{L}^t_1 = (1,1,9)$,
$\mathbf{L}^t_2 = (0,2,6)$ and $\mathbf{L}^t_3 = (0,0,30)$.

There is a second way of describing a finite lattice. Have a
parallelepiped defined by three {\sl compact edge vectors},
$\mathbf{L}^c_i$, such that not one of the six distinct face
diagonals joining a pair of corner vertices will have a length
less than any of the compact edge vectors. These compact vectors
are easily obtained by linear matrix multiplying the utlf matrix.
The edge vectors, $\mathbf{L}^c_1$, $\mathbf{L}^c_2$ and
$\mathbf{L}^c_3$, of the above example of N = 15 vertices of the
parallelepiped are the set of compact defining vectors for this
finite bcc lattice.
\section{Generation of finite ferromagnetic bcc
lattices} \label{generation} From equations (\ref{primvects}) and
(\ref{edgevects}) it is clear that, in each of the three vectors
defining a finite bcc lattice, each of its components are odd or
each are even. For a {\em bipartite} finite bcc lattice, all
components of each of the defining vectors are even. In a recent
paper \cite{Betts98} on the S = 1/2 Heisenberg antiferromagnet
using exact diagonalization on bipartite lattices only (of
course), we found a total of forty useful bipartite bcc lattices
with $16 \le \mathrm{N} \le 32$ vertices. The defining vectors are
listed in Table 2 of that paper. The smallest bcc bipartite
lattice of use has N = 16 vertices so that each vertex would have
a complete set of nearest neighbours on the other sublattice. The
large diagonalization time on the computer meant that bipartite
lattices of N $>$ 32 would not be useful.

In this paper on the S = 1/2 XY ferromagnet exact diagonalization
can be used on all finite bcc lattices. Furthermore we felt that
finite lattices of as few as N = 9 vertices would sometimes be
suitable as each vertex would have a complete set of eight
neighbours. Nevertheless more often finite bcc lattices need to
have $\mathrm{N} \ge 15$ so that each vertex would have a complete
set of second neighbours also.

Here we show in Table 1 that for 9 $\le \mathrm{N} \le$ 25 there
is a total of 72 odd-N bcc lattices, for 10 $\le \mathrm{N} \le$
26 there is a total of 125 even-N bcc lattices. Only 13 of these
125 bcc lattices are bipartite and therefore usable for exact
diagonalization of Hamiltonians of antiferromagnetic models.
However, for 16 $\le \mathrm{N} \le$ 32 there are 40 bipartite bcc
lattices - enough to use in the statistical estimates of the
properties of the S = 1/2 Heisenberg antiferromagnet at T = 0.

In Table 1 we also show that for 9 $\le \mathrm{N} \le$ 32 there
are 156 odd-N bcc lattices and 306 even-N bcc lattices, of which
181 have N = 28, 30 or 32 vertices. For computation time purpose
we decided not to include so many more even-N bcc lattices. In the
last double column, the numbers of the finite bcc lattices most
useful for the exact diagonalization method are listed.

The geometric criterion we used to get rid of a minority of poor
lattices is that the geometric imperfection, $\mathrm{I_G}$, of
each lattice divided by N should be less than 0.35.
\begin{table}
\caption{Numbers of distinct finite odd/even bcc lattices and
useful distinct lattices of $9 \le \mathrm{N} \le 32$ vertices. n
is the total number of lattices, $\mathrm{n_b}$ is the number of
bipartite lattices and $\nu$ is the number of useful lattices for
each value of N.}
\label{numlattices}       
\centering\vspace{2mm}
\begin{tabular}{ccccccc}
\hline\noalign{\smallskip}
 N & \hspace{0.5cm} & n & \hspace{0.5cm} & $\mathrm{n_b}$ & \hspace{0.5cm} & $\nu$ \\
\noalign{\smallskip}\hline\noalign{\smallskip}
 9/10 & & 1/1 & & 0 & & 1/1 \\
 11/12 & & 1/2 & & 0 & & 1/2 \\
 13/14 & & 2/3 & & 0 & & 2/3 \\
 15/16 & & 7/6 & & 1 & & 7/6 \\
 17/18 & & 5/13 & & 1 & & 4/12 \\
 19/20 & & 7/18 & & 2 & & 6/16 \\
 21/22 & & 17/15 & & 1 & & 16/13 \\
 23/24 & & 12/42 & & 6 & & 11/36 \\
 25/26 & & 20/25 & & 2 & & 14/19 \\
 \cline{1-1} \cline{3-3} \cline{5-5} \cline{7-7} Subtotals & & 72/125 & & 13 & & 62/108 \\
 \cline{1-1} \cline{3-3} \cline{5-5} \cline{7-7} 27/28 & & 36/45 & & 5 & & 25/32 \\
 29/30 & & 22/67 & & 7 & & 16/44 \\
 31/32 & & 26/69 & & 15 & & 17/50 \\
\cline{1-1} \cline{3-3} \cline{5-5} \cline{7-7} Totals & & 156/306
& & 40 & & 120/234 \\ \noalign{\smallskip}\hline
\end{tabular}
\end{table}
Geometric imperfection, for finite simple cubic lattices, was
first introduced by Betts and Stewart \cite{BettsStewart97}. In an
{\em infinite} bcc lattice each vertex has geometrically
$\mathrm{N}_1 = 8$ nearest neighbours, $\mathrm{N}_2 = 6$ second
nearest neighbours, $\mathrm{N}_3 = 12$ third, $\mathrm{N}_4 = 24$
fourth, $\mathrm{N}_5 = 8$ fifth, ... $\mathrm{N}_k$ of $k^{th}$
neighbours. We can call these "shells" of neighbours about any
vertex. In a {\em finite} lattice with an {\em odd} number, N, of
vertices one vertex, labelled A, is chosen arbitrarily and
temporarily as a centre or origin. Then the remaining N-1 vertices
are automatically paired, say B \& C, D \& E, ... in such a way
that, under inversion in A, the other vertices interchange B
$\leftrightarrow$ C, D $\leftrightarrow$ E, etc.

Now consider a {\em finite} lattice of N vertices in which all
shells below the k shell are full. If the k shell is not full it
will have $\mathrm{N}_k - 2$, $\mathrm{N}_k - 4$ or ... 0
vertices. But if all shells above it are empty the finite lattice
has geometric perfection; $\mathrm{I_G} = 0$. However, if the k +
1 shell above it has a pair of vertices and all shells above that
are empty, then $\mathrm{I_G} = 2$.

The geometric imperfection of 4 can occur in two ways: a) if shell
$k-1$ has $\mathrm{N}_{k-1}-2$ vertices, the shell $k$ has
$\mathrm{N}_k$ vertices and the shell $k+1$ has 2 vertices then
$\mathrm{I_G} = 4$, or b) if shell $k$ has $\mathrm{N}_{k} - 4$
vertices and the last occupied shell, $k+1$, has 4 vertices,
$\mathrm{I_G} = 4$.

Let us again use the example of N = 15 lattice 15.28. Any vertex
of this lattice has 8 nearest neighbours, only 4 second nearest
neighbours and 2 third nearest neighbours. Thus the geometric
imperfection $\mathrm{I_G}(15.28) = 2$. Another N = 15 lattice,
15.577, has 8 nearest neighbours and 6 second nearest neighbours
of any vertex. So this lattice is a perfect N = 15 bcc lattice
with $\mathrm{I_G}(15.577) = 0$.

Determining the geometric imperfection of an even N lattice is
just a bit different. Again one vertex is chosen as the temporary
origin and N-2 vertices are thus paired so that under inversion in
the origin, A, the pairs of other vertices exchange B
$\leftrightarrow$ C, D $\leftrightarrow$ E, ... . However, the
remaining vertex, Z, stays where it is under inversion in A. Thus
A and Z are temporarily the poles. Because pole Z may be farther
from the origin than in a perfect finite lattice, it means that
geometric imperfections in even N lattices can be odd or even. The
simplest example among finite bcc lattices is an N = 12 vertex
lattice labeled 12.17. It has defining vectors in compact form
$\mathbf{L}^c_1 = (1,-1,3)$, $\mathbf{L}^c_2 = (-1,3,1)$ and
$\mathbf{L}^c_3 = (4,2,0)$. There is a complete set of four pairs
of nearest neighbours, one pair of second nearest neighbours and a
single third neighbour, so $\mathrm{I_G}(12.17) = 1$. On the other
hand, the twelve vertex bcc lattice 12.25 has four pairs of
nearest neighbours, one pair of second neighbours and a single
second neighbour to any vertex. Thus $\mathrm{I_G}(12.25) = 0$.

We have a computer program that can obtain, in utlf form, all
finite lattices of any number of vertices, N, on any infinite
bipartite lattice in two or three dimensions. The program can also
yield for each finite lattice the geometrical imperfection,
$\mathrm{I_G}$, the topological imperfection, $\mathrm{I_T}$, the
rotational symmetry in Schoenfliess notation, S, and whether or
not the finite lattice is bipartite. ( The topological
imperfection is not used in this article, but it was described in
our previous paper \cite{Betts98} ). If we were studying a
Heisenberg antiferromagnet only bipartite finite lattices could be
used. However, as we have been studying an S = 1/2 XY
$ferromagnet$ on the infinite bcc lattice we could and did use
both bipartite and nonbipartite finite bcc lattices of even N and
also finite bcc lattices of odd N vertices.

After considerable computing and statistical analysis we decided
that only those finite bcc lattices with $\mathrm{I_G/N} < 0.35$
should be used. The finite bcc lattices must have at least N = 9
vertices, otherwise no vertex would have a complete set of eight
nearest neighbours. We stopped at bcc lattices of N = 32 because
diagonalization of Hamiltonians on lattices with more than 32
vertices would require an excessive amount of computer memory and
especially time. The finite bcc lattices now suitable are
described in Tables A1, A2 and A3. Notice that there are 116
ferromagnetically useful odd-N bcc lattices of $15 \le \mathrm{N}
\le 31$ and 102 even-N bcc lattices of $16 \le \mathrm{N} \le 26$.
For N = 28, 30 and 32 there are 27 bipartite bcc lattices, because
we decided 245 finite bcc lattices would be sufficient. After all,
in the earlier paper \cite{Betts98} a total of 40 bipartite
lattices of $16 \le \mathrm{N} \le 32$ seemed sufficient. There
are another 99 nonbipartite bcc lattices of N = 28, 30 and 32.

\section{Computation of physical properties of the S = 1/2 XY
ferromagnet on finite bcc lattices} \label{computation} The
Hamiltonian of the spin one-half XY ferromagnet in zero magnetic
field is
\begin{equation}
\label{hamiltonian} H = J \sum_{<i,j>} (S_i^x S_j^x + S_i^y S_j^y)
\end{equation}
where the sum is over nearest neighbour pairs of vertices. It was
proved by Lieb and Mattis \cite{Lieb62} that the ground state of
this model on an infinite three-dimensional lattice has total spin
component in the z direction equal to zero and is nondegenerate.
The first excited state has the z component of the total spin
equal to one or minus one. Later it was proved by Kennedy et al
\cite{Kennedy882} and Kubo and Kishi \cite{Kubo88} that this model
has long range order, $m_{\perp}^2 = \hspace{1em} <m_x^2 + m_y^2>
\hspace{0.5em} = 2<m_x^2>$, in the ground state (and obviously for
$\mathrm{T} < \mathrm{T_c}$).

All of the finite bcc lattices are translationally invariant and
invariant under inversion, so this simplifies the diagonalization
of the Hamiltonians. On the even-N bcc lattices the ground state
Hamiltonian submatrix has the z component of the total spin
$\mathrm{S}_z = 0$. The first excited state submatrix has
$\mathrm{S}_z = \pm 1$. On the odd-N bcc lattices the ground state
has the z component of the total spin $\mathrm{S}_z = \pm 1/2$.
The first excited state has $\mathrm{S}_z = \pm 3/2$. In the
Hamiltonian diagonalizations we use the submatrices with positive
$\mathrm{S}_z$'s.

\sloppy On each of the submatrices of the Hamiltonian the ground
state energy and its corresponding eigenvectors have been
calculated. The Lanczos technique used in the diagonalization is
standard \cite{Cullum32}. On the even-N bcc lattices we computed
on the $\mathrm{S_z}$ = 0 submatrix the ground state energy
$\mathrm{E}_0 = \mathrm{N}\epsilon_0$ and on the $\mathrm{S_z}$ =
1 submatrix the first excited state energy, $\mathrm{E}_1 =
\mathrm{N}\epsilon_1$. Using the ground state and first excited
state eigenvectors we computed the square of the magnetizations in
the spin-space X direction, $<m_x^2>$. Hence we immediately
obtained on each even-N bcc lattice $m_{x,0}$ and $m_{x,1}$
respectively. The computations have been made on all even-N
lattices of $10 \le \mathrm{N} \le 26$ vertices, and on bipartite
lattices only for $28 \le \mathrm{N} \le 32$. On all odd-N bcc
lattices of $9 \le \mathrm{N} \le 31$ we computed on the
$\mathrm{S_z} = 1/2$ Hamiltonian submatrix the ground state energy
$\mathrm{E}_{1/2} = \mathrm{N} \epsilon_{1/2}$ and on the
$\mathrm{S_z}$ = 3/2 submatrix the first excited state energy
$\mathrm{E}_{3/2} = \mathrm{N} \epsilon_{3/2}$.

\fussy If any two geometrically distinct finite bcc lattices have
the same ground state and first excited state energies then they
are topologically identical, and only one of them would be used.
The computed results are listed in Tables B1 and B2 for all finite
bcc lattices used.

In Table B1 we show the dimensionless energies per vertex,
$\epsilon_s$, for the $\mathrm{S_z}$ = 1/2 ground state, the
lowest excited $\mathrm{S_z}$ = 3/2 state energy and the ground
state magnetization squared on all odd-N bcc lattices of $9 \le
\mathrm{N} \le 31$, for which $\mathrm{I_G/N} < 0.35$. The total
number of such bcc lattices is 120; of these odd-N lattices 33
have more than 28 vertices.

In Table B2 we also show $\epsilon_s$ for the $\mathrm{S_z}$ = 0
ground state, the lowest $\mathrm{S_z}$ = 1 state energy and the
ground state magnetization squared on all even-N bcc lattices of
$10 \le \mathrm{N} \le 26$ and for which $\mathrm{I_G/N} < 0.35$.
There is a total of 108 such lattices. Among these lattices only
13 are bipartite and hence suitable for exact diagonalization of
the Hamiltonians of {\em antiferromagnets} such as the S = 1/2
Heisenberg antiferromagnet \cite{Betts98}.

Of even-N bipartite lattices with N = 28, 30 or 32 for which
$\mathrm{I_G/N} < 0.35$ there are only 27. However, there are 99
nonbipartite bcc lattices of $28 \le \mathrm{N} \le 32$. It would
thus have taken, for these lattices, several thousands of hours of
computer time to diagonalize all the $\mathrm{S}_z = 0$ and
$\mathrm{S}_z = 1$ submatrix Hamiltonians of the S = 1/2 XY
ferromagnet. Thus we use only the bipartite bcc lattices in this
range.

Statistical analysis of these data will be described in the next
section. The purpose is to derive the T = 0 physical properties on
the infinite bcc lattice.

\section{Statistical analysis estimation of the T = 0 properties
of the S = 1/2 XY ferromagnet on the infinite body-centred cubic
lattice} \label{statistical} In order to estimate any physical
properties of the S = 1/2 XY ferromagnet on the infinite bcc
lattice, using the exact diagonalization data for that property on
the finite bcc lattices, one must fit the data statistically to
the appropriate finite lattice scaling equation. The independent
variable in such equations is $\mathrm{L}^{-1}$ defined by
$\mathrm{L}^3 = \mathrm{N}$, the number of vertices. For example,
the energy of the XY ferromagnet finite lattice scaling equation
for any total spin $\mathrm{S_z}$ (or for simplicity below, $s$)
is in three dimensions
\begin{equation}
\label{energyscale} \epsilon_s(\mathrm{L}) = \epsilon_s(\infty) +
\mathrm{A}^s_4\mathrm{L}^{-4} + \mathrm{A}^s_6\mathrm{L}^{-6} +
...
\end{equation}
Such scaling equations are explained carefully in the paper by
Hasenfratz and Niedermayer \cite{Hasenfratz93}, and earlier papers
on this matter \cite{Hasenfratz90,Neuberger89,Fisher89} are cited
therein. Oitmaa et al \cite{Oitmaa94} used both series expansions
and spin wave methods on the S = 1/2 Heisenberg antiferromagnet on
the simple cubic and bcc lattices to obtain estimates of the first
term and the second term of equation (\ref{energyscale}). They
obtained estimates of the ground state energy per vertex,
$\epsilon_0(\infty)$ from the first term and spin wave velocity,
v, from the second term.

In the pioneering application of the method of exact
diagonalization of the finite lattices of the S = 1/2 XY
ferromagnet on simple cubic lattices \cite{BettsStewart97} it was
quickly noticed that the curve fitting the $\epsilon_{1/2}$ data
from odd-N simple cubic lattices was definitely above the curve
fitting the $\epsilon_0$ data from even-N lattices. Moreover, here
we also have data for $\epsilon_1$ on even-N and $\epsilon_{3/2}$
on odd-N bcc lattices. It is clear that while on finite lattices
$\epsilon_1 > \epsilon_0$ and $\epsilon_{3/2} > \epsilon_{1/2}$,
on the {\em infinite} lattices one should expect $\epsilon_0 =
\epsilon_{1/2} = \epsilon_{1} = \epsilon_{3/2}$. Similarly, we
should expect $\mathrm{A}_4^0 = \mathrm{A}_4^{1/2} =
\mathrm{A}_4^1 = \mathrm{A}_4^{3/2}$ on the infinite lattices in
three dimensions.

Clearly we do not study the energies, $E_s$, of the S = 1/2 XY
ferromagnet (or other models) but the dimensionless energies per
vertex, $\epsilon_s = E_s / \mathrm{N}J$. Nevertheless, on finite
lattices $\epsilon_s$ depends on both the number of vertices, N,
and the geometric arrangement of those N vertices on these finite
bcc lattices so our energy density label is
$\epsilon_s(\mathrm{N.i})$.

\begin{figure}
\begin{center}
\resizebox{0.66\textwidth}{!}{ 
 \includegraphics[angle=-90]{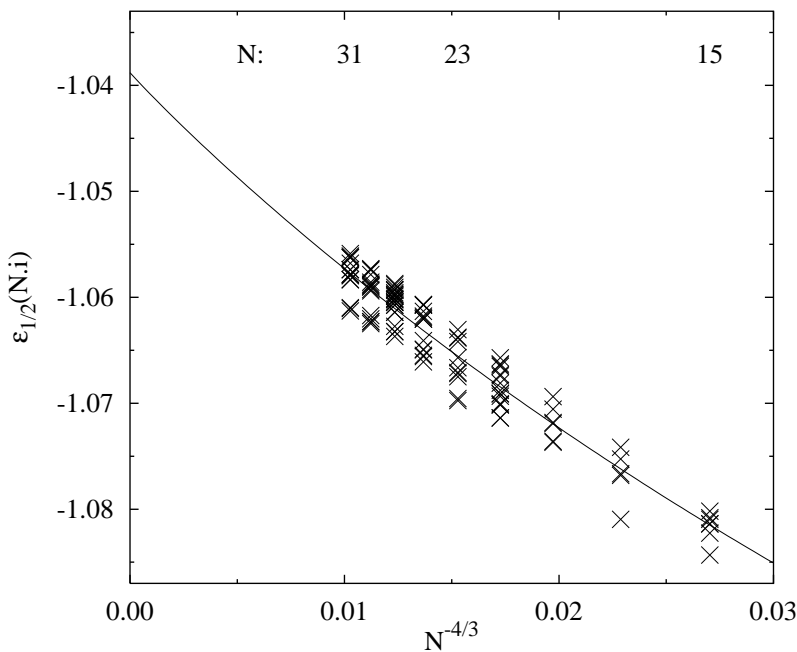}
 }
\end{center}
\caption{Above each horizontal position at $\mathrm{N}^{-4/3}$ for
N = 15, 17, 19, ... 31 is a set of points representing the bcc XY
ferromagnet energies per vertex, $\epsilon_{1/2}(\mathrm{N.i})$,
of each of the finite bcc lattices of N vertices. The statistical
curve fitted to these points is $\epsilon_{1/2}(\mathrm{N} =
\epsilon(\infty) + \mathrm{A}_4^{1/2} \mathrm{N}^{-4/3} +
\mathrm{A}_6^{1/2} \mathrm{N}^{-2}$.}
\label{fig2a}       
\end{figure}

Our statistical analyses have used sets of $\epsilon_s$(N.i) data
for $\mathrm{N} \ge 15$, which means that in each finite bcc
lattice each vertex can have a complete shell of nearest
neighbours and a complete shell of second nearest neighbours. As
an example of the scatter of the ground state energy per vertex,
$\epsilon_{1/2}(\mathrm{N.i})$, of the S = 1/2 XY ferromagnet in
each of nine sets of vertices $15 \le \mathrm{N} \le 31$ is
displayed in Figure 2. Notice that a small number of the points in
each of the nine sets of clusters are "outriders" relatively far
from the centre of the cluster and are not statistically useful.
Statistical analysis of these data determines the numerical
coefficients of equation (7) and hence the curve in Fig. 2. We
also exclude data from finite lattices for which the geometric
imperfection, $\mathrm{I_G} \ge 0.35$. Clearly the smaller sets of
data are better because the statistical estimates of
$\epsilon_s(\infty)$ and of $\mathrm{A}_4^s$ are considerably
closer to one another when N $\ge 15$ than they are when
$\mathrm{N} \ge 9$ data are used. Accordingly in Table 2 we
display only the statistical estimates obtained using data of
$\mathrm{N} \ge 15$ and $\mathrm{I_G} < 0.35$.

\begin{table}
\caption{Statistically analysed coefficients of energy equation
(\ref{energyscale}) using ground states or first excited states on
even-N or odd-N bcc lattices. B means bipartite and Nb means
nonbipartite.}
\label{energyfits}       
\centering\vspace{2mm}
\begin{tabular}{r@{\hspace{0.1cm}$\le$\hspace{0.2cm}}c@{\hspace{0.1cm}$\le$\hspace{0.2cm}}lcc@{\hspace{0.2cm}}c@{\hspace{0.2cm}}c}
\hline\noalign{\smallskip} \multicolumn{3}{c}{Data sets} &
$\mathrm{S_z}$ & $\epsilon_s (\infty)$ & $\mathrm{A}_4^s$ &
$\mathrm{A}_6^s$ \\ \noalign{\smallskip}\hline\noalign{\smallskip}
 16 & N & 26, Nb & 0 & -1.0393(1) & -2.22(3) & 2.8 \\
 16 & N & 32, B & 0 & -1.0418(5) & -1.71(3) & 1.3 \\
 15 & N & 31, Nb & 1/2 & -1.0388(1) & -2.27(4) & 4.2 \\
 16 & N & 26, Nb & 1 & -1.0391(3) & -2.26(5) & 7.5 \\
 16 & N & 32, B & 1 & -1.0417(2) & -1.75(5) & 6.0 \\
 15 & N & 31, Nb & 3/2 & -1.0385(4) & -2.35(11) & 13.4 \\
  \noalign{\smallskip}\hline
\end{tabular}
\end{table}

In Table \ref{energyfits} we have displayed $\epsilon_s(\infty)$,
$\mathrm{A}_4^s$ and $\mathrm{A}_6^s$, the coefficients in
equation (\ref{energyscale}) for $s = 0$, 1/2, 1 and 3/2. Indeed
for $s = 0$ and $s = 1$ we have obtained separate statistical
estimates using only bipartite lattices (B) and only nonbipartite
lattices (Nb). (Of course, all finite bcc lattices of odd N must
be nonbipartite). Notice that to four digits
$\epsilon_s^{Nb}(\infty) = -1.039$ for each of the four Nb sets of
$s = 0$, 1/2, 1 and 3/2. On the other hand $\epsilon_0^B(\infty) =
\epsilon_1^B(\infty) = -1.042$. The averages are respectively
$\epsilon^{Nb}(\infty) = -1.0389$ and $\epsilon^B(\infty) =
-1.0417$. We must conclude that the $\epsilon^{Nb}(\infty)$
average estimate, using some two hundred nonbipartite lattices, is
more accurate than $\epsilon^B(\infty)$ using only forty bipartite
lattices.

As is known \cite{Hasenfratz93}, the second coefficient,
$\mathrm{A}_4^s$, of the finite lattice scaling equation is
independent of $s$. In Table \ref{energyfits} the average of the
four nonbipartite estimates is $\mathrm{A}_4^{Nb} = -2.27$,
considerably larger than the average $\mathrm{A}_4^B = -1.73$. It
is also known that $\mathrm{A}_6^s$ does increase substantially as
$s$ increases, and our Table \ref{energyfits} shows the
accelerating increase of $\mathrm{A}_6^s$ from $s$ = 0 to $s$ =
3/2.

According to Oitmaa et al \cite{Oitmaa94} and others
\cite{Hasenfratz93}, \cite{Hasenfratz90}, \cite{Neuberger89}, on
the bcc lattice $-\mathrm{A}_4 = \beta$v/2 where the geometric
shape factor $\beta = 2.1104607$ and v is the spin wave velocity.
Using the average estimate of $\mathrm{A}_4$ we obtain $\mathrm{v}
= 2.15(2)$. (Oitmaa et al \cite{Oitmaa94} give $\beta$ for the bcc
lattice as different by a factor of $2^{4/3}$ from what we show
here. This is due to the fact that they use $\mathrm{L}^3$ = N/2
where we use $\mathrm{L}^3$ = N.)

Because we have used approximately two hundred and sixty finite
lattices of $15 \le \mathrm{N} \le 32$ vertices, we would fully
clutter any figure with so many points. However, we can insert a
much smaller set of points (together with curves) in our figures
and thus avoid clutter. For each N there is a subset of exact
diagonalization values; the averages only of these physical
properties provide points in the appropriate figure. Such a figure
will also contain one or more curves representing the finite
lattice scaling equation determined statistically by using {\em
all} useful data not just the averages at each N.

\begin{figure}
\begin{center}
\resizebox{0.66\textwidth}{!}{ 
 \includegraphics[angle=-90]{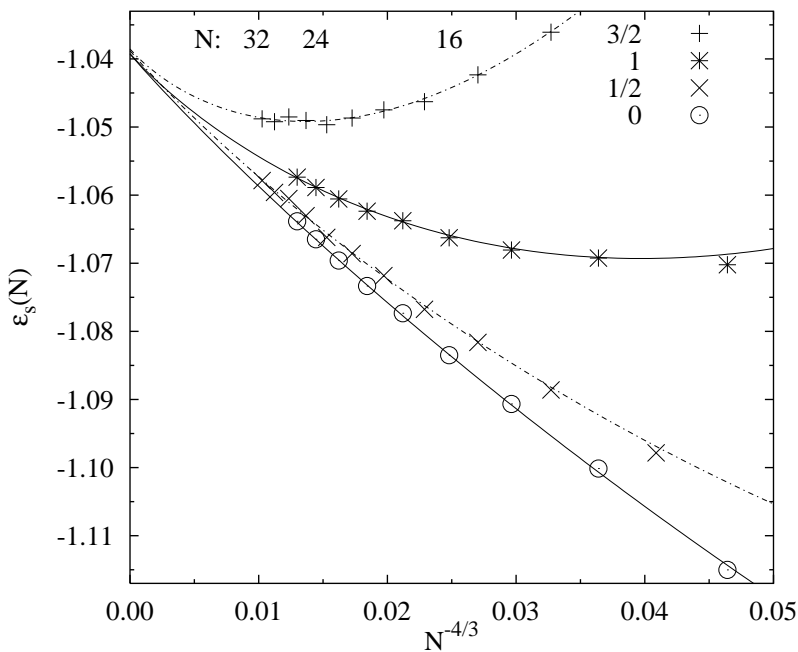}
 }
\end{center}
\caption{Energy per vertex, $\epsilon$, against $\mathrm{L}^{-4}$
for $\mathrm{S_z}$ = 3/2, 1, 1/2 and 0. The points are the
averages, $\epsilon_s(\mathrm{N})$, for each set of $\mathrm{S_z}$
and N. The curves are from equation (\ref{energyscale}).}
\label{engL4}       
\end{figure}

Figure \ref{engL4} is an example with four curves representing
scaling equation (\ref{energyscale}), $\mathrm{S_z} \equiv s$ = 0,
1/2, 1 and 3/2, and about 40 averaged points,
$\epsilon_s(\mathrm{N}) =
\hspace{0.25em}\langle\epsilon_s(\mathrm{N.i})\rangle$. Which
points with which $\mathrm{S_z}$ is shown in the upper right Fig.
\ref{engL4}. For even N we have used only the data for
nonbipartite lattices.

All the curves in Fig. \ref{engL4} reach the same point,
$\epsilon(\infty)$, on the vertical axis. Furthermore, each of the
four curves has the same slope, $\mathrm{A}_4$ at 1/$\mathrm{L}^4$
= 0. However, the curvature increases rapidly with $\mathrm{S_z}$.
These effects demonstrate the data in Table \ref{energyfits}.

For each of three reasons it is certain that $\mathrm{A}_6^s$ is
not independent of $s$. First, from Table 2 it is clear that
$\mathrm{A}_6^0 < \mathrm{A}_6^{1/2} < \mathrm{A}_6^1 <
\mathrm{A}_6^{3/2}$. Second, the finite lattice scaling equation
curves in Fig. \ref{engL4} have a greater curvature the larger
$\mathrm{S_z}$ is. Finally, Hasenfratz and Niedermayer
\cite{Hasenfratz93} show that
\begin{equation}
\label{hnengdiff} [\epsilon_1(\mathrm{L}) -
\epsilon_0(\mathrm{L})] = \mathrm{D}_6 \mathrm{L}^{-6} +
\mathrm{D}_8 \mathrm{L}^{-8} + ...
\end{equation}
In three dimensions this means that $\epsilon_1(\infty) -
\epsilon_0(\infty) = 0$ and that $\mathrm{A}_4^1 - \mathrm{A}_4^0
= 0$ but $\mathrm{A}_6^1 - \mathrm{A}_6^0 = \mathrm{D}_6 \neq 0$.
Further, $\mathrm{D}_6 = 1/2\chi_{\perp}$, the inverse
susceptibility. We define $\Delta = \epsilon_1(\mathrm{L}) -
\epsilon_0(\mathrm{L})$ and $\Delta' = \epsilon_{3/2} -
\epsilon_{1/2}$. Then our statistical analysis of even-N bcc S =
1/2 XY ferromagnet data obeys the scaling equation
\begin{equation}
\label{Ldeltascale}
 \mathrm{L}^6 \Delta = \mathrm{D}_6 + \mathrm{D}_8 \mathrm{L}^{-2} + ...
\end{equation}
with the same form for odd-N.

This shows that $\mathrm{D}_0 = \mathrm{D}_1 = ... \hspace{1em}
\mathrm{D}_5 = \mathrm{D}_7 = 0$. These data also show that
$\mathrm{D}_6$ and $\mathrm{D}_8$ are not equal to zero. After
observing graphs and examining Table 3 we believe that the
two-term scaling equations are best.

\begin{table}
\caption{Statistically analysed coefficients of bcc finite lattice
data of equation (9) where $\mathrm{D}_n$ are coefficients of even
N data and $\mathrm{D}_n'$ are equations for odd N.}
\label{deltafits}       
\centering\vspace{2mm}
\begin{tabular}{@{}lccc@{}}
\hline\noalign{\smallskip}
 Coefficients and & \multicolumn{3}{c}{\underline{Number of terms in the scaling equation}} \\
 their ratios & one term & two terms & three terms \\ \noalign{\smallskip}\hline\noalign{\smallskip}
 $\mathrm{D}_6$ & 4.4119 & 4.2531 & 4.2800 \\
 $\mathrm{D}_6'$ & 8.7844 & 8.5087 & 8.5654 \\
 $\mathrm{D}_6' / \mathrm{D}_6$ & 1.9911 & 2.0006 & 2.0013 \\
 $\mathrm{D}_8$ & ---- & 1.022 & 0.624 \\
 $\mathrm{D}_8'$ & ---- & 1.995 & 1.150 \\
 $\mathrm{D}_8' / \mathrm{D}_8$ & ---- & 1.95 & 1.84 \\
\noalign{\smallskip}\hline
\end{tabular}
\end{table}

Based on the average of the three of our ratios in Table 3,
$\mathrm{D}_6' / \mathrm{D}_6 = 1.998(1)$, clearly $\mathrm{D}_6'
= 2 \mathrm{D}_6$. The averages in the two ratios $\mathrm{D}_8' /
\mathrm{D}_8 = 1.90(2)$ indicate also that $\mathrm{D}_8' \sim 2
\mathrm{D}_8$. These factors of 2 arise from equation (2.19) in
the paper of Hasenfratz and Neidermayer \cite{Hasenfratz93}.

The susceptibility, $\chi_{\perp}$, is defined as $\chi_{\perp} =
1 / 2\mathrm{D}_6 = 1/\mathrm{D}_6'$ according to Hasenfratz and
Neidermayer \cite{Hasenfratz93}. Using Table 3 these coefficients
in the column of two terms the estimates of $\chi_{\perp}$ are
0.1176 and 0.1175 so we take the average $\chi_{\perp} =
0.1176(4)$. Without calculating the spin stiffness, $\rho_s$,
directly, but using the data above, $\rho_s = \mathrm{v}^2
\chi_{\perp} = 0.55(4)$.

Now we turn to the magnetization. Oitmaa et al \cite{Oitmaa94},
following Neuberger and Ziman \cite{Neuberger89}, determined by
effective Lagrangian theory that the second term in the finite
lattice scaling equation for the staggered magnetization is
proportional to $\mathrm{L}^{d-1}$. Thus for the Heisenberg {\em
antiferromagnet} in three dimensions the finite lattice scaling
equation for the {\em staggered magnetization} is
\begin{equation}
\label{smagscale} m^+(\mathrm{L}) = m^+(\infty) + \mathrm{B}_2
\mathrm{L}^{-2} + ...
\end{equation}
In our paper \cite{Betts98} on the S = 1/2 Heisenberg
antiferromagnet on the bcc lattices we found via our statistical
analyses that
\begin{equation}
\label{pmagscale} m^+(\mathrm{L}) = m^+(\infty) + \mathrm{B}_2
\mathrm{L}^{-2} + \mathrm{B}_4 \mathrm{L}^{-4} + ...
\end{equation}

The XY ferromagnet on bipartite lattices has a magnetization,
$m_{\perp}(\mathrm{L})$, that is identical to the XY
antiferromagnet's $m^+$. Thus we expect the same finite lattice
scaling equation for $m_{\perp}(\mathrm{L})$ on the set of all bcc
lattices. The data that we list in Appendix B is the magnetization
squared in the X spin space direction. We decided later that it
was more convenient to deal with $m_{\perp}$ so the data were
converted according to $m_{\perp} = \sqrt{2 m_x^2}$.

Our statistical analyses of the magnetization of the S = 1/2 XY
ferromagnet on bcc lattices found that the finite lattice scaling
equation for the magnetization, $m_{\perp}$, is indeed of the same
form as (\ref{pmagscale}). (Our guess is that a fourth term might
well be $\mathrm{B}_6 / \mathrm{L}^6$.)

\begin{table}[ht!]
\caption{Statistical results of fitting equation (\ref{pmagscale})
to the finite lattice data for $m_{\perp,s}$. The Odd-N and
Even-N, bipartite data were taken in the range from $15 \le
\mathrm{N} \le 32$. The Even-N, nonbipartite data were in the
range $16 \le \mathrm{N} \le 26$.}
\label{magest}       
\centering\vspace{2mm}
\begin{tabular}{l@{\hspace{0.1cm}}c@{\hspace{0.1cm}}ccc}
\hline\noalign{\smallskip} Data set of lattices & $\mathrm{S_z}$ &
$m_{\perp}(\infty)$ & $\mathrm{B}_2$ & $\mathrm{B}_4$ \\
\noalign{\smallskip}\hline\noalign{\smallskip}
 Odd-N & 1/2 & 0.478(1) & 0.31(2) & -0.03 \\
 Even-N, bipartite & 0 & 0.478(1) & 0.27(3) & 0.2 \\
 Even-N, nonbipartite & 0 & 0.485(1) & 0.19(3) & 0.5 \\
 \noalign{\smallskip}\hline
\end{tabular}
\end{table}

Table \ref{magest} shows that the estimates of $m_{\perp}(\infty)$
and $\mathrm{B}_2$ using the bipartite even-N lattices are closer
to the estimates of the same properties using the odd-N lattices
than those using the nonbipartite even-N lattices. This may be due
to the fact that the latter data stop at N = 26 while the first
two go to 31 and 32 respectively.

Equation (52) in Oitmaa et al \cite{Oitmaa94} implies that
\begin{equation}
\label{oitmaaB2rho} \mathrm{B}_2= \Sigma\mathrm{v}\gamma / \rho_s
\end{equation}
for the Heisenberg antiferromagnet in three dimensions. $\Sigma$
is their $m^+({\infty})$ and $\gamma$ is another shape factor. As
$m_{\perp}$ scales the same as $\mathrm{m}^+$, there is no reason
why (\ref{oitmaaB2rho}) shouldn't apply for $m_{\perp}$ and we
obtain a means to check our earlier result for $\rho_s$. In our
notation, (\ref{oitmaaB2rho}) becomes
\begin{equation}
\label{ourB2rho} \rho_s = m_{\perp}(\infty) \mathrm{v} \gamma /
\mathrm{B}_2
\end{equation}
where $m_{\perp}(\infty) = 0.480$, $\mathrm{B}_2 = 0.29$
(averaging from the top two estimates of Table \ref{magest}),
$\gamma = 0.1792055$ (from \cite{Oitmaa94}) and v = 2.16, our
earlier result (again, our value of $\gamma$ is a factor of
$2^{2/3}$ different from that given in \cite{Oitmaa94}, for the
same reason as stated before concerning $\beta$). This gives us a
second estimate of $\rho_s = 0.64$. This agrees quite well with
our earlier estimate of $\rho_s = 0.55$ given the indirect route
of each estimate.

After statistically analysing the magnetization data to estimate
the coefficients $m_{\perp}(\infty)$, $\mathrm{B}_2$ and
$\mathrm{B}_4$ in finite lattice scaling equation
(\ref{pmagscale}) the question arises as to whether some of these
coefficients, like the coefficients $\epsilon(\infty)$ and
$\mathrm{A}_4$ in (\ref{energyscale}), are independent of the spin
state of the data. Compared with the energy per vertex, there is
not the same physical importance to study $m_{\perp}$ finite
lattice data above the ground states, $\mathrm{S_z}$ = 0 for
even-N and $\mathrm{S_z}$ = 1/2 for odd-N lattices. Nevertheless,
after studying the $\mathrm{L}^6 (\epsilon_s - \epsilon_{s-1})$
finite lattice data we decided to study $\mathrm{L}^n (m_{\perp,0}
- m_{\perp,1})$ data. For $m_{\perp,1}$ we had data only for
bipartite bcc lattices of $16 \le \mathrm{N} \le 26$. We soon
found that $n$ = 6 was the appropriate exponent of L above, and
that in the finite lattice scaling equation (\ref{pmagscale})
$m_{\perp,1}(\infty) = m_{\perp,0}(\infty)$, $\mathrm{B}_2^1 =
\mathrm{B}_2^0$ and $\mathrm{B}_4^1 = \mathrm{B}_4^0$, as well as
our small amount of data showed. We then used a scaling equation
analogous to (\ref{Ldeltascale}), namely
\begin{equation}
\label{magdeltascale}
 \mathrm{L}^6 \Gamma \equiv (m_{\perp,0} -
 m_{\perp,1}) \mathrm{L}^6 = \mathrm{F}_6 + \mathrm{F}_8 \mathrm{L}^{-2} + ...
\end{equation}
We found that with two terms in (\ref{magdeltascale})
\begin{equation}
 \mathrm{L}^6 \Gamma = 0.942 - 0.17 \mathrm{L}^{-2}
\end{equation}
If it were of greater physical interest we could readily compute
many $m_{\perp,3/2}$ data and more $m_{\perp,1}$ data especially
on nonbipartite even-N finite bcc lattices.

\section{Summary, conclusions and outlook}
Following our earlier paper \cite{Betts98} on the S = 1/2 {\em
Heisenberg antiferromagnet} on finite even-N {\em bipartite}
lattices, we have here extended the generation of finite bcc
lattices to include all {\em nonbipartite} lattices, of $9 \le
\mathrm{N} \le 31$ odd lattices and of $10 \le \mathrm{N} \le 32$
even lattices. We have found it useful statistically to delete the
small fraction of these lattices for which the geometric
imperfection, $\mathrm{I_G}$, is greater than or equal to 0.35 N
and to delete also lattices of fewer than N = 15 vertices.

On each of the remaining even-N bcc lattices of $10 \le \mathrm{N}
\le 26$ we computed, in the lowest state of $\mathrm{S_z}$ = 0 and
of $\mathrm{S_z}$ = 1, of the S = 1/2 XY ferromagnet Hamiltonian
the dimensionless energy per vertex, $\epsilon_0(\mathrm{N.i})$
and $\epsilon_1(\mathrm{N.i})$, and the dimensionless
magnetization squared per vertex, $m_{\perp,0}^2(\mathrm{N.i})$.
For N = 28, 30 and 32 we used only the bipartite lattices because
there are 99 nonbipartite even-N bcc lattices in this range.
Similarly, on the odd-N bcc lattices of $9 \le \mathrm{N} \le 31$
we computed $\epsilon_{1/2}(\mathrm{N.i})$,
$\epsilon_{3/2}(\mathrm{N.i})$ and
$m_{\perp,1/2}^2(\mathrm{N.i})$. Because it appeared of little
physical significance we computed $m_{\perp,1}^2$ on only the
bipartite lattices of $16 \le \mathrm{N} \le 26$.

We used the same type of statistical analyses of the data
described above as we had used in our previous paper
\cite{Betts98}. The energy data statistically fitted to the
appropriate three-term finite lattice scaling equation provided
some interesting results. First Table 2 shows that within
confidence limits $\epsilon_s(\infty)$ is independent of the spin
$\mathrm{S_z}$ for at least $\mathrm{S_z}$ = 0, 1/2, 1 and 3/2.
Furthermore, $\mathrm{A}_4^s$, the coefficient of the second term
in (\ref{energyscale}), $\mathrm{A}_4^s \mathrm{L}^{-4}$, is also
independent of $\mathrm{S_z}$ within confidence limits. However,
$\mathrm{A}_6^s$, the third term coefficient is decidedly
dependent on $\mathrm{S_z}$. Our findings for
$\epsilon_s(\infty)$, $\mathrm{A}_4^s$ and $\mathrm{A}_6^s$
correspond to the general rules for finite size scaling equations
as shown by Hasenfratz and Niedermayer \cite{Hasenfratz93}. The
spin wave velocity, v, is determined by $\mathrm{A}_4$, and the
susceptibility, $\chi_{\perp}$, is determined by $\mathrm{D}_6$.
Hence the spin stiffness, $\rho_s$, is also estimated.

Turning to the magnetization per vertex, $m_{\perp,s}$, we
calculated only the data for $m_{\perp,0}(\mathrm{N.i})$ and
$m_{\perp,1/2}(\mathrm{N.i})$ for N $\le 32$. There are two
reasons for not obtaining such data for $S_z ( = s ) > 1/2$.
First, because it is obvious and well known that
$m_{\perp,0}(\infty) = m_{\perp,s}(\infty)$. Second, because there
is no known physical property analogous to the susceptibility
obtained by studying $\epsilon_{s+1}(\mathrm{L}) -
\epsilon_s(\mathrm{L})$ as $\mathrm{L} \rightarrow \infty$ as
discussed on p. 10.

At this point we compare our estimates of $\epsilon_0$ and
$m_{\perp}$ with spin wave and series expansion results of Oitmaa
et al \cite{Oitmaa94}. Unfortunately, they have not included
additional terms in the scaling equations.
\begin{table}[ht!]
\caption{Comparison of our exact diagonalization estimates of
ground state energy and magnetization with estimates by other
methods.}
\label{results}       
\centering\vspace{2mm}
\begin{tabular}{lcc}
\hline\noalign{\smallskip}
 Method of calculation & $\epsilon(\infty)$ & $m_{\perp}(\infty)$ \\
 \noalign{\smallskip}\hline\noalign{\smallskip}
 Exact diagonalization, all lattices & -1.0396 & 0.4803 \\
 Exact diagonalization, nonbipartite & -1.0389 & 0.4816 \\
 Spin wave, first order & -1.0380 & 0.4829 \\
 Spin wave, second order & -1.0406 & 0.4822 \\
 Series expansion, seven terms & -1.0408 & 0.4824 \\
 \noalign{\smallskip}\hline
\end{tabular}
\end{table}
Our average of exact diagonalization estimates of
$\epsilon_s(\infty)$ is within 1\% of the results of second order
spin wave and a seven term series expansion. Our average estimate
of $m_{\perp,s}(\infty)$ is within 2\% of second order spin wave
and series expansion results. All three methods could be used
further to obtain more precise estimates of energy, magnetization,
etc..

\sloppy Powerful computers are evolving rapidly with memory
getting larger, calculations getting faster, etc. These advances
will prove particularly advantageous for the method of exact
diagonalization of Hamiltonians on finite lattices at zero
temperature vis-a-vis other methods such as spin wave. This exact
diagonalization method is especially useful in three dimensions at
zero temperature where some other methods may not work so well.
Soon the exact diagonalization method will be a feasible way to
study at high precision models with three states per lattice
vertex - spin 1 antiferromagnets, the t-J model, Hubbard model,
etc. Of course, the exact diagonalization study of models with two
states per vertex will obtain still more precise estimates of the
physical properties at zero temperature.

\fussy The finite three-dimensional lattices used so far are the
three simplest: simple cubic, bcc and fcc lattices with complete
cubic symmetry and only one vertex per unit cell. As computers
advance various models on other lattices of lower symmetry and/or
more vertices per unit cell could be studied. For example,
perovskite lattices, for many maganites, have eight vertices per
unit cell. The exact diagonalization method on finite lattices
could eventually work on simple models of two states per vertex
but at finite temperature. Much more can be just over the horizon.

\section*{Acknowledgements}
We have received useful points from Prof. C.J. Hamer. This
research has been supported in part by the Natural Sciences and
Engineering Research Council of Canada, the Imperial Oil
Charitable Foundation, and by Deutsche Forschungsgemeinschaft
(project Ri 6151/1-2).

\newpage
\appendix
\section*{Appendix A: Geometric properties of finite bcc lattices}
Each different bcc lattice is labelled N.i where N is its number
of vertices and i is the index distinguishing lattices of the same
N.

\noindent {\bf Table A1.} Finite bcc lattices with an odd number
of vertices, $9 \le \mathrm{N} \le 31$

\vspace{2mm}
\begin{tabular}{r@{.}lcccc}
\hline\noalign{\smallskip} N & i & $\mathbf{L}^t_1$ &
$\mathbf{L}^t_2$ & $\mathbf{L}^t_3$ & $\mathrm{I_G}$ \\
\noalign{\smallskip}\hline\noalign{\smallskip}
 9 & 13 & ( 1, 1, 7) & ( 0, 2, 4) & ( 0, 0,18) & 0 \\
 11 & 15 & ( 1, 1, 7) & ( 0, 2, 4) & ( 0, 0,22) & 0 \\
 13 & 17 & ( 1, 1, 7) & ( 0, 2, 4) & ( 0, 0,26) & 2 \\
 13 & 18 & ( 1, 1, 9) & ( 0, 2, 4) & ( 0, 0,26) & 0 \\
 15 & 19 & ( 1, 1, 7) & ( 0, 2, 4) & ( 0, 0,30) & 4 \\
 15 & 20 & ( 1, 1, 9) & ( 0, 2, 4) & ( 0, 0,30) & 2 \\
 15 & 21 & ( 1, 1,11) & ( 0, 2, 4) & ( 0, 0,30) & 2 \\
 15 & 28 & ( 1, 1, 9) & ( 0, 2, 6) & ( 0, 0,30) & 2 \\
 15 & 30 & ( 1, 1,13) & ( 0, 2, 6) & ( 0, 0,30) & 2 \\
 15 & 38 & ( 1, 1,13) & ( 0, 2, 8) & ( 0, 0,30) & 2 \\
 15 & 577 & ( 1, 3, 3) & ( 0, 6, 0) & ( 0, 0,10) & 0 \\
 17 & 22 & ( 1, 1, 9) & ( 0, 2, 4) & ( 0, 0,34) & 2 \\
 17 & 23 & ( 1, 1,11) & ( 0, 2, 4) & ( 0, 0,34) & 0 \\
 17 & 24 & ( 1, 1,13) & ( 0, 2, 4) & ( 0, 0,34) & 2 \\
 17 & 25 & ( 1, 1,15) & ( 0, 2, 6) & ( 0, 0,34) & 2 \\
 19 & 24 & ( 1, 1, 9) & ( 0, 2, 4) & ( 0, 0,38) & 2 \\
 19 & 25 & ( 1, 1,11) & ( 0, 2, 4) & ( 0, 0,38) & 0 \\
 19 & 26 & ( 1, 1,13) & ( 0, 2, 4) & ( 0, 0,38) & 0 \\
 19 & 27 & ( 1, 1,15) & ( 0, 2, 4) & ( 0, 0,38) & 4 \\
 19 & 37 & ( 1, 1,15) & ( 0, 2, 6) & ( 0, 0,38) & 2 \\
 19 & 38 & ( 1, 1,17) & ( 0, 2, 6) & ( 0, 0,38) & 4 \\
 21 & 26 & ( 1, 1, 9) & ( 0, 2, 4) & ( 0, 0,42) & 4 \\
 21 & 27 & ( 1, 1,11) & ( 0, 2, 4) & ( 0, 0,42) & 0 \\
 21 & 28 & ( 1, 1,13) & ( 0, 2, 4) & ( 0, 0,42) & 0 \\
 21 & 29 & ( 1, 1,15) & ( 0, 2, 4) & ( 0, 0,42) & 4 \\
 21 & 30 & ( 1, 1,17) & ( 0, 2, 4) & ( 0, 0,42) & 4 \\
 21 & 37 & ( 1, 1, 9) & ( 0, 2, 6) & ( 0, 0,42) & 6 \\
 21 & 38 & ( 1, 1,11) & ( 0, 2, 6) & ( 0, 0,42) & 2 \\
 21 & 40 & ( 1, 1,15) & ( 0, 2, 6) & ( 0, 0,42) & 0 \\
 21 & 41 & ( 1, 1,17) & ( 0, 2, 6) & ( 0, 0,42) & 0 \\
 21 & 42 & ( 1, 1,19) & ( 0, 2, 6) & ( 0, 0,42) & 0 \\
 21 & 49 & ( 1, 1,11) & ( 0, 2, 8) & ( 0, 0,42) & 4 \\
 21 & 53 & ( 1, 1,19) & ( 0, 2, 8) & ( 0, 0,42) & 4 \\
 21 & 73 & ( 1, 1,15) & ( 0, 2,12) & ( 0, 0,42) & 4 \\
 21 & 75 & ( 1, 1,19) & ( 0, 2,12) & ( 0, 0,42) & 0 \\
 21 & 86 & ( 1, 1,19) & ( 0, 2,14) & ( 0, 0,42) & 4 \\
 21 & 1332 & ( 1, 3, 3) & ( 0, 6, 0) & ( 0, 0,14) & 0 \\
 23 & 28 & ( 1, 1, 9) & ( 0, 2, 4) & ( 0, 0,46) & 6 \\
 23 & 29 & ( 1, 1,11) & ( 0, 2, 4) & ( 0, 0,46) & 0 \\
 23 & 30 & ( 1, 1,13) & ( 0, 2, 4) & ( 0, 0,46) & 2 \\
 23 & 31 & ( 1, 1,15) & ( 0, 2, 4) & ( 0, 0,46) & 2 \\
 23 & 32 & ( 1, 1,17) & ( 0, 2, 4) & ( 0, 0,46) & 8 \\
 23 & 33 & ( 1, 1,19) & ( 0, 2, 4) & ( 0, 0,46) & 2 \\
 23 & 41 & ( 1, 1,11) & ( 0, 2, 6) & ( 0, 0,46) & 4 \\
 23 & 45 & ( 1, 1,19) & ( 0, 2, 6) & ( 0, 0,46) & 6 \\
 23 & 46 & ( 1, 1,21) & ( 0, 2, 6) & ( 0, 0,46) & 2 \\
 23 & 53 & ( 1, 1,11) & ( 0, 2, 8) & ( 0, 0,46) & 6 \\
 \noalign{\smallskip}\hline
\end{tabular}
\hspace{1mm}
\begin{tabular}{r@{.}lc@{\hspace{0.3cm}}c@{\hspace{0.3cm}}cc}
 \hline\noalign{\smallskip} N & i & $\mathbf{L}^t_1$ &
 $\mathbf{L}^t_2$ & $\mathbf{L}^t_3$ & $\mathrm{I_G}$ \\
 \noalign{\smallskip}\hline\noalign{\smallskip}
 23 & 56 & ( 1, 1,17) & ( 0, 2, 8) & ( 0, 0,46) & 6 \\
 25 & 30 & ( 1, 1, 9) & ( 0, 2, 4) & ( 0, 0,50) & 8 \\
 25 & 31 & ( 1, 1,11) & ( 0, 2, 4) & ( 0, 0,50) & 2 \\
 25 & 32 & ( 1, 1,13) & ( 0, 2, 4) & ( 0, 0,50) & 2 \\
 25 & 33 & ( 1, 1,15) & ( 0, 2, 4) & ( 0, 0,50) & 4 \\
 25 & 34 & ( 1, 1,17) & ( 0, 2, 4) & ( 0, 0,50) & 4 \\
 25 & 36 & ( 1, 1,21) & ( 0, 2, 4) & ( 0, 0,50) & 2 \\
  25 & 44 & ( 1, 1,11) & ( 0, 2, 6) & ( 0, 0,50) & 6 \\
  25 & 45 & ( 1, 1,13) & ( 0, 2, 6) & ( 0, 0,50) & 6 \\
  25 & 49 & ( 1, 1,21) & ( 0, 2, 6) & ( 0, 0,50) & 8 \\
  25 & 50 & ( 1, 1,23) & ( 0, 2, 6) & ( 0, 0,50) & 4 \\
  25 & 58 & ( 1, 1,13) & ( 0, 2, 8) & ( 0, 0,50) & 2 \\
  25 & 59 & ( 1, 1,15) & ( 0, 2, 8) & ( 0, 0,50) & 8 \\
  25 & 72 & ( 1, 1,15) & ( 0, 2,10) & ( 0, 0,50) & 2 \\
  25 & 4227 & ( 1, 3, 5) & ( 0,10, 0) & ( 0, 0,10) & 2 \\
  27 & 33 & ( 1, 1,11) & ( 0, 2, 4) & ( 0, 0,54) & 4 \\
  27 & 34 & ( 1, 1,13) & ( 0, 2, 4) & ( 0, 0,54) & 2 \\
  27 & 35 & ( 1, 1,15) & ( 0, 2, 4) & ( 0, 0,54) & 6 \\
  27 & 36 & ( 1, 1,17) & ( 0, 2, 4) & ( 0, 0,54) & 4 \\
  27 & 39 & ( 1, 1,23) & ( 0, 2, 4) & ( 0, 0,54) & 2 \\
  27 & 49 & ( 1, 1,15) & ( 0, 2, 6) & ( 0, 0,54) & 4 \\
  27 & 51 & ( 1, 1,19) & ( 0, 2, 6) & ( 0, 0,54) & 2 \\
  27 & 52 & ( 1, 1,21) & ( 0, 2, 6) & ( 0, 0,54) & 4 \\
  27 & 53 & ( 1, 1,23) & ( 0, 2, 6) & ( 0, 0,54) & 4 \\
  27 & 54 & ( 1, 1,25) & ( 0, 2, 6) & ( 0, 0,54) & 4 \\
  27 & 62 & ( 1, 1,13) & ( 0, 2, 8) & ( 0, 0,54) & 4 \\
  27 & 66 & ( 1, 1,21) & ( 0, 2, 8) & ( 0, 0,54) & 8 \\
  27 & 68 & ( 1, 1,25) & ( 0, 2, 8) & ( 0, 0,54) & 2 \\
  27 & 81 & ( 1, 1,23) & ( 0, 2,10) & ( 0, 0,54) & 4 \\
  27 & 91 & ( 1, 1,15) & ( 0, 2,12) & ( 0, 0,54) & 8 \\
  27 & 94 & ( 1, 1,21) & ( 0, 2,12) & ( 0, 0,54) & 6 \\
  27 & 124 & ( 1, 1,25) & ( 0, 2,16) & ( 0, 0,54) & 4 \\
  27 & 136 & ( 1, 1,21) & ( 0, 2,18) & ( 0, 0,54) & 8 \\
  27 & 2354 & ( 1, 1, 5) & ( 0, 6, 0) & ( 0, 0,18) & 2 \\
  27 & 2396 & ( 1, 1, 5) & ( 0, 6, 6) & ( 0, 0,18) & 0 \\
  27 & 2399 & ( 1, 1,11) & ( 0, 6, 6) & ( 0, 0,18) & 4 \\
  27 & 2549 & ( 1, 3, 3) & ( 0, 6, 0) & ( 0, 0,18) & 4 \\
  27 & 2550 & ( 1, 3, 5) & ( 0, 6, 0) & ( 0, 0,18) & 2 \\
  27 & 2594 & ( 1, 3, 9) & ( 0, 6, 6) & ( 0, 0,18) & 2 \\
  27 & 14309 & ( 3, 3, 3) & ( 0, 6, 0) & ( 0, 0, 6) & 0 \\
  29 & 35 & ( 1, 1,11) & ( 0, 2, 4) & ( 0, 0,58) & 4 \\
  29 & 36 & ( 1, 1,13) & ( 0, 2, 4) & ( 0, 0,58) & 2 \\
  29 & 37 & ( 1, 1,15) & ( 0, 2, 4) & ( 0, 0,58) & 4 \\
  29 & 38 & ( 1, 1,17) & ( 0, 2, 4) & ( 0, 0,58) & 6 \\
  29 & 39 & ( 1, 1,19) & ( 0, 2, 4) & ( 0, 0,58) & 4 \\
  29 & 41 & ( 1, 1,23) & ( 0, 2, 4) & ( 0, 0,58) & 8 \\
 \noalign{\smallskip}\hline
\end{tabular}
\begin{tabular}{r@{.}lc@{\hspace{0.3cm}}c@{\hspace{0.3cm}}cc}
 \hline\noalign{\smallskip} N & i & $\mathbf{L}^t_1$ &
 $\mathbf{L}^t_2$ & $\mathbf{L}^t_3$ & $\mathrm{I_G}$ \\
 \noalign{\smallskip}\hline\noalign{\smallskip}
  29 & 42 & ( 1, 1,25) & ( 0, 2, 4) & ( 0, 0,58) & 2 \\
  29 & 50 & ( 1, 1,11) & ( 0, 2, 6) & ( 0, 0,58) & 10 \\
  29 & 51 & ( 1, 1,13) & ( 0, 2, 6) & ( 0, 0,58) & 8 \\
  29 & 52 & ( 1, 1,15) & ( 0, 2, 6) & ( 0, 0,58) & 2 \\
  29 & 55 & ( 1, 1,21) & ( 0, 2, 6) & ( 0, 0,58) & 8 \\
  29 & 57 & ( 1, 1,25) & ( 0, 2, 6) & ( 0, 0,58) & 4 \\
  29 & 58 & ( 1, 1,27) & ( 0, 2, 6) & ( 0, 0,58) & 4 \\
  29 & 66 & ( 1, 1,13) & ( 0, 2, 8) & ( 0, 0,58) & 4 \\
  29 & 73 & ( 1, 1,27) & ( 0, 2, 8) & ( 0, 0,58) & 4 \\
  29 & 85 & ( 1, 1,21) & ( 0, 2,10) & ( 0, 0,58) & 8 \\
  31 & 37 & ( 1, 1,11) & ( 0, 2, 4) & ( 0, 0,62) & 4 \\
  31 & 38 & ( 1, 1,13) & ( 0, 2, 4) & ( 0, 0,62) & 2 \\
  31 & 39 & ( 1, 1,15) & ( 0, 2, 4) & ( 0, 0,62) & 2 \\
  31 & 40 & ( 1, 1,17) & ( 0, 2, 4) & ( 0, 0,62) & 8 \\
  \noalign{\smallskip}\hline
\end{tabular}
\hspace{2mm}
\begin{tabular}{r@{.}lc@{\hspace{0.3cm}}c@{\hspace{0.3cm}}cc}
 \hline\noalign{\smallskip} N & i & $\mathbf{L}^t_1$ &
 $\mathbf{L}^t_2$ & $\mathbf{L}^t_3$ & $\mathrm{I_G}$ \\
 \noalign{\smallskip}\hline\noalign{\smallskip}
  31 & 41 & ( 1, 1,19) & ( 0, 2, 4) & ( 0, 0,62) & 4 \\
  31 & 44 & ( 1, 1,25) & ( 0, 2, 4) & ( 0, 0,62) & 4 \\
  31 & 45 & ( 1, 1,27) & ( 0, 2, 4) & ( 0, 0,62) & 2 \\
 31 & 54 & ( 1, 1,13) & ( 0, 2, 6) & ( 0, 0,62) & 10 \\
 31 & 55 & ( 1, 1,15) & ( 0, 2, 6) & ( 0, 0,62) & 2 \\
 31 & 58 & ( 1, 1,21) & ( 0, 2, 6) & ( 0, 0,62) & 2 \\
 31 & 59 & ( 1, 1,23) & ( 0, 2, 6) & ( 0, 0,62) & 8 \\
 31 & 61 & ( 1, 1,27) & ( 0, 2, 6) & ( 0, 0,62) & 6 \\
 31 & 62 & ( 1, 1,29) & ( 0, 2, 6) & ( 0, 0,62) & 4 \\
 31 & 71 & ( 1, 1,15) & ( 0, 2, 8) & ( 0, 0,62) & 8 \\
 31 & 74 & ( 1, 1,21) & ( 0, 2, 8) & ( 0, 0,62) & 4 \\
 31 & 78 & ( 1, 1,29) & ( 0, 2, 8) & ( 0, 0,62) & 4 \\
 31 & 94 & ( 1, 1,29) & ( 0, 2,10) & ( 0, 0,62) & 2 \\
\noalign{\smallskip}\hline
\end{tabular}

\vspace{8mm}
\noindent {\bf Table A2.} Finite bcc lattices with an even number
of lattices, $10 \le \mathrm{N} \le 26$

\vspace{2mm}
\begin{tabular}{r@{.}lc@{\hspace{0.3cm}}c@{\hspace{0.3cm}}cc}
\hline\noalign{\smallskip} N & i & $\mathbf{L}^t_1$ &
$\mathbf{L}^t_2$ & $\mathbf{L}^t_3$ & $\mathrm{I_G}$ \\
 \noalign{\smallskip}\hline\noalign{\smallskip}
 10 & 15 & ( 1, 1, 7) & ( 0, 2, 4) & ( 0, 0,20) & 0 \\
 12 & 25 & ( 1, 1, 9) & ( 0, 2, 6) & ( 0, 0,24) & 0 \\
 12 & 17 & ( 1, 1, 7) & ( 0, 2, 4) & ( 0, 0,24) & 1 \\
 14 & 21 & ( 1, 1,11) & ( 0, 2, 4) & ( 0, 0,28) & 1 \\
 14 & 20 & ( 1, 1, 9) & ( 0, 2, 4) & ( 0, 0,28) & 2 \\
 14 & 19 & ( 1, 1, 7) & ( 0, 2, 4) & ( 0, 0,28) & 3 \\
 16 & 23 & ( 1, 1,11) & ( 0, 2, 4) & ( 0, 0,32) & 0 \\
 16 & 1642 & ( 2, 0, 8) & ( 0, 2, 4) & ( 0, 0,16) & 1 \\
 16 & 41 & ( 1, 1,11) & ( 0, 2, 8) & ( 0, 0,32) & 2 \\
 16 & 24 & ( 1, 1,13) & ( 0, 2, 4) & ( 0, 0,32) & 3 \\
 16 & 32 & ( 1, 1,11) & ( 0, 2, 6) & ( 0, 0,32) & 3 \\
 16 & 21 & ( 1, 1, 7) & ( 0, 2, 4) & ( 0, 0,32) & 5 \\
 18 & 37 & ( 1, 1,15) & ( 0, 2, 6) & ( 0, 0,36) & 0 \\
 18 & 25 & ( 1, 1,11) & ( 0, 2, 4) & ( 0, 0,36) & 1 \\
 18 & 1101 & ( 1, 3, 3) & ( 0, 6, 0) & ( 0, 0,12) & 1 \\
 18 & 2024 & ( 2, 0, 8) & ( 0, 2, 4) & ( 0, 0,18) & 1 \\
 18 & 24 & ( 1, 1, 9) & ( 0, 2, 4) & ( 0, 0,36) & 2 \\
 18 & 26 & ( 1, 1,13) & ( 0, 2, 4) & ( 0, 0,36) & 2 \\
 18 & 35 & ( 1, 1,11) & ( 0, 2, 6) & ( 0, 0,36) & 2 \\
 18 & 36 & ( 1, 1,13) & ( 0, 2, 6) & ( 0, 0,36) & 2 \\
 18 & 1001 & ( 1, 1, 3) & ( 0, 6, 0) & ( 0, 0,12) & 2 \\
 18 & 27 & ( 1, 1,15) & ( 0, 2, 4) & ( 0, 0,36) & 3 \\
 18 & 34 & ( 1, 1, 9) & ( 0, 2, 6) & ( 0, 0,36) & 4 \\
 18 & 67 & ( 1, 1,15) & ( 0, 2,12) & ( 0, 0,36) & 5 \\
 20 & 27 & ( 1, 1,11) & ( 0, 2, 4) & ( 0, 0,40) & 0 \\
 20 & 50 & ( 1,1,13) & ( 0, 2, 8) & ( 0, 0,40) & 0 \\
 20 & 30 & ( 1,1,17) & ( 0, 2, 4) & ( 0, 0,40) & 1 \\
 20 & 62 & ( 1, 1,15) & ( 0, 2,10) & ( 0, 0,40) & 1 \\
 \noalign{\smallskip}\hline
\end{tabular}
\hspace{1mm}
\begin{tabular}{r@{.}lc@{\hspace{0.3cm}}c@{\hspace{0.3cm}}cc}
\hline\noalign{\smallskip} N & i & $\mathbf{L}^t_1$ &
$\mathbf{L}^t_2$ & $\mathbf{L}^t_3$ & $\mathrm{I_G}$ \\
\noalign{\smallskip}\hline\noalign{\smallskip}
 20 & 630 & ( 1, 1, 7) & ( 0, 4, 4) & ( 0, 0,20) & 2 \\
 20 & 2446 & ( 2, 0, 8) & ( 0, 2, 4) & ( 0, 0,20) & 2 \\
 20 & 26 & ( 1, 1, 9) & ( 0, 2, 4) & ( 0, 0,40) & 3 \\
 20 & 38 & ( 1, 1,11) & ( 0, 2, 6) & ( 0, 0,40) & 3 \\
 20 & 61 & ( 1, 1,13) & ( 0, 2,10) & ( 0, 0,40) & 3 \\
 20 & 2447 & ( 2, 0,10) & ( 0, 2, 4) & ( 0, 0,20) & 3 \\
 20 & 73 & ( 1, 1,15) & ( 0, 2,12) & ( 0, 0,40) & 4 \\
 20 & 629 & ( 1, 1, 5) & ( 0, 4, 4) & ( 0, 0,20) & 4 \\
 20 & 29 & ( 1, 1,15) & ( 0, 2, 4) & ( 0, 0,40) & 5 \\
 20 & 40 & ( 1, 1,15) & ( 0, 2, 6) & ( 0, 0,40) & 5 \\
 20 & 63 & ( 1, 1,17) & ( 0, 2,10) & ( 0, 0,40) & 5 \\
 20 & 51 & ( 1, 1,15) & ( 0, 2, 8) & ( 0, 0,40) & 6 \\
 22 & 29 & ( 1, 1,11) & ( 0, 2, 4) & ( 0, 0,44) & 0 \\
 22 & 33 & ( 1, 1,19) & ( 0, 2, 4) & ( 0, 0,44) & 1 \\
 22 & 31 & ( 1, 1,15) & ( 0, 2, 4) & ( 0, 0,44) & 2 \\
 22 & 43 & ( 1, 1,15) & ( 0, 2, 6) & ( 0, 0,44) & 2 \\
 22 & 44 & ( 1, 1,17) & ( 0, 2, 6) & ( 0, 0,44) & 4 \\
 22 & 30 & ( 1, 1,13) & ( 0, 2, 4) & ( 0, 0,44) & 3 \\
 22 & 2908 & ( 2, 0, 8) & ( 0, 2, 4) & ( 0, 0,22) & 3 \\
 22 & 41 & ( 1, 1,11) & ( 0, 2, 6) & ( 0, 0,44) & 4 \\
 22 & 28 & ( 1, 1, 9) & ( 0, 2, 4) & ( 0, 0,44) & 5 \\
 22 & 66 & ( 1, 1,13) & ( 0, 2,10) & ( 0, 0,44) & 5 \\
 22 & 69 & ( 1, 1,19) & ( 0, 2,10) & ( 0, 0,44) & 5 \\
 22 & 42 & ( 1, 1,13) & ( 0, 2, 6) & ( 0, 0,44) & 6 \\
 22 & 32 & ( 1, 1,17) & ( 0, 2, 4) & ( 0, 0,44) & 7 \\
 24 & 30 & ( 1, 1, 9) & ( 0, 2, 4) & ( 0, 0,48) & 7 \\
 24 & 31 & ( 1, 1,11) & ( 0, 2, 4) & ( 0, 0,48) & 1 \\
 \noalign{\smallskip}\hline
\end{tabular}

\vspace{2mm}
\begin{tabular}{r@{.}lc@{\hspace{0.3cm}}c@{\hspace{0.3cm}}cc}
\hline\noalign{\smallskip} N & i & $\mathbf{L}^t_1$ &
$\mathbf{L}^t_2$ & $\mathbf{L}^t_3$ & $\mathrm{I_G}$ \\
\noalign{\smallskip}\hline\noalign{\smallskip}
 24 & 32 & ( 1, 1,13) & ( 0, 2, 4) & ( 0, 0,48) & 2 \\
 24 & 33 & ( 1, 1,15) & ( 0, 2, 4) & ( 0, 0,48) & 2 \\
 24 & 34 & ( 1, 1,17) & ( 0, 2, 4) & ( 0, 0,48) & 8 \\
 24 & 36 & ( 1, 1,21) & ( 0, 2, 4) & ( 0, 0,48) & 1 \\
 24 & 44 & ( 1, 1,11) & ( 0, 2, 6) & ( 0, 0,48) & 6 \\
 24 & 45 & ( 1, 1,13) & ( 0, 2, 6) & ( 0, 0,48) & 8 \\
 24 & 46 & ( 1, 1,15) & ( 0, 2, 6) & ( 0, 0,48) & 5 \\
 24 & 47 & ( 1, 1,17) & ( 0, 2, 6) & ( 0, 0,48) & 1 \\
 24 & 48 & ( 1, 1,19) & ( 0, 2, 6) & ( 0, 0,48) & 1 \\
 24 & 49 & ( 1, 1,21) & ( 0, 2, 6) & ( 0, 0,48) & 1 \\
 24 & 57 & ( 1, 1,11) & ( 0, 2, 8) & ( 0, 0,48) & 7 \\
 24 & 58 & ( 1, 1,13) & ( 0, 2, 8) & ( 0, 0,48) & 1 \\
 24 & 59 & ( 1, 1,15) & ( 0, 2, 8) & ( 0, 0,48) & 6 \\
 24 & 61 & ( 1, 1,19) & ( 0, 2, 8) & ( 0, 0,48) & 8 \\
 24 & 71 & ( 1, 1,13) & ( 0, 2,10) & ( 0, 0,48) & 5 \\
 24 & 73 & ( 1, 1,17) & ( 0, 2,10) & ( 0, 0,48) & 5 \\
 24 & 86 & ( 1, 1,17) & ( 0, 2,12) & ( 0, 0,48) & 2 \\
 24 & 87 & ( 1, 1,19) & ( 0, 2,12) & ( 0, 0,48) & 4 \\
 24 & 88 & ( 1, 1,21) & ( 0, 2,12) & ( 0, 0,48) & 3 \\
 24 & 100 & ( 1, 1,19) & ( 0, 2,14) & ( 0, 0,48) & 3 \\
 24 & 113 & ( 1, 1,19) & ( 0, 2,16) & ( 0, 0,48) & 6 \\
 24 & 873 & ( 1, 1, 5) & ( 0, 4, 4) & ( 0, 0,24) & 6 \\
 24 & 875 & ( 1, 1, 9) & ( 0, 4, 4) & ( 0, 0,24) & 6 \\
 24 & 888 & ( 1, 1, 9) & ( 0, 4, 6) & ( 0, 0,24) & 2 \\
 24 & 894 & ( 1, 1,21) & ( 0, 4, 6) & ( 0, 0,24) & 6 \\
 24 & 1860 & ( 1, 3, 3) & ( 0, 6, 0) & ( 0, 0,16) & 2 \\
 24 & 1911 & ( 1, 3, 1) & ( 0, 6, 8) & ( 0, 0,16) & 6 \\
\noalign{\smallskip}\hline
\end{tabular}
\hspace{2mm}
\begin{tabular}{r@{.}lc@{\hspace{0.3cm}}c@{\hspace{0.3cm}}cc}
\hline\noalign{\smallskip} N & i & $\mathbf{L}^t_1$ &
$\mathbf{L}^t_2$ & $\mathbf{L}^t_3$ & $\mathrm{I_G}$ \\
\noalign{\smallskip}\hline\noalign{\smallskip}
 24 & 2705 & ( 1, 3, 3) & ( 0, 8, 0) & ( 0, 0,12) & 3 \\
 24 & 3410 & ( 2, 0, 8) & ( 0, 2, 4) & ( 0, 0,24) & 4 \\
 24 & 3411 & ( 2, 0,10) & ( 0, 2, 4) & ( 0, 0,24) & 4 \\
 24 & 3424 & ( 2, 0,10) & ( 0, 2, 6) & ( 0, 0,24) & 4 \\
 24 & 4266 & ( 2, 0, 4) & ( 0, 4, 6) & ( 0, 0,12) & 4 \\
 24 & 3412 & ( 2, 0,12) & ( 0, 2, 4) & ( 0, 0,24) & 5 \\
 24 & 3425 & ( 2, 0,12) & ( 0, 2, 6) & ( 0, 0,24) & 5 \\
 26 & 33 & ( 1, 1,11) & ( 0, 2, 4) & ( 0, 0,52) & 3 \\
 26 & 34 & ( 1, 1,13) & ( 0, 2, 4) & ( 0, 0,52) & 2 \\
 26 & 35 & ( 1, 1,15) & ( 0, 2, 4) & ( 0, 0,52) & 7 \\
 26 & 36 & ( 1, 1,17) & ( 0, 2, 4) & ( 0, 0,52) & 4 \\
 26 & 38 & ( 1, 1,21) & ( 0, 2, 4) & ( 0, 0,52) & 5 \\
 26 & 39 & ( 1, 1,23) & ( 0, 2, 4) & ( 0, 0,52) & 3 \\
 26 & 47 & ( 1, 1,11) & ( 0, 2, 6) & ( 0, 0,52) & 7 \\
 26 & 48 & ( 1, 1,13) & ( 0, 2, 6) & ( 0, 0,52) & 8 \\
 26 & 49 & ( 1, 1,15) & ( 0, 2, 6) & ( 0, 0,52) & 4 \\
 26 & 50 & ( 1, 1,17) & ( 0, 2, 6) & ( 0, 0,52) & 4 \\
 26 & 51 & ( 1, 1,19) & ( 0, 2, 6) & ( 0, 0,52) & 8 \\
 26 & 52 & ( 1, 1,21) & ( 0, 2, 6) & ( 0, 0,52) & 7 \\
 26 & 61 & ( 1, 1,11) & ( 0, 2, 8) & ( 0, 0,52) & 9 \\
 26 & 76 & ( 1, 1,13) & ( 0, 2,10) & ( 0, 0,52) & 8 \\
 26 & 78 & ( 1, 1,17) & ( 0, 2,10) & ( 0, 0,52) & 4 \\
 26 & 92 & ( 1, 1,17) & ( 0, 2,12) & ( 0, 0,52) & 3 \\
 26 & 122 & ( 1, 1,21) & ( 0, 2,16) & ( 0, 0,52) & 5 \\
 26 & 5912 & ( 2, 0, 8) & ( 0, 2, 4) & ( 0, 0,26) & 5 \\
 26 & 5913 & ( 2, 0,10) & ( 0, 2, 4) & ( 0, 0,26) & 5 \\
 \noalign{\smallskip}\hline
\end{tabular}

\vspace{8mm}
\noindent {\bf Table A3.} Bipartite finite lattices with $28 \le
\mathrm{N} \le 32$

\vspace{2mm}
\begin{tabular}{r@{.}lc@{\hspace{0.3cm}}c@{\hspace{0.3cm}}cc}
\hline\noalign{\smallskip} N & i & $\mathbf{L}^t_1$ &
$\mathbf{L}^t_2$ & $\mathbf{L}^t_3$ & $\mathrm{I_G}$ \\
\noalign{\smallskip}\hline\noalign{\smallskip}
 28 & 6785 & ( 2, 0,10) & ( 0, 2, 4) & ( 0, 0,28) & 5 \\
 28 & 6786 & ( 2, 0,12) & ( 0, 2, 4) & ( 0, 0,28) & 5 \\
 28 & 6816 & ( 2, 0,12) & ( 0, 2, 8) & ( 0, 0,28) & 5 \\
 28 & 6784 & ( 2, 0, 8) & ( 0, 2, 4) & ( 0, 0,28) & 7 \\
 28 & 6800 & ( 2, 0,10) & ( 0, 2, 6) & ( 0, 0,28) & 7 \\
 30 & 7717 & ( 2, 0,10) & ( 0, 2, 4) & ( 0, 0,30) & 4 \\
 30 & 7718 & ( 2, 0,12) & ( 0, 2, 4) & ( 0, 0,30) & 4 \\
 30 & 7733 & ( 2, 0,10) & ( 0, 2, 6) & ( 0, 0,30) & 4 \\
 30 & 7734 & ( 2, 0,12) & ( 0, 2, 6) & ( 0, 0,30) & 4 \\
 30 & 7750 & ( 2, 0,12) & ( 0, 2, 8) & ( 0, 0,30) & 4 \\
 30 & 7716 & ( 2, 0, 8) & ( 0, 2, 4) & ( 0, 0,30) & 8 \\
 30 & 10754 & ( 2, 0, 4) & ( 0, 6, 0) & ( 0, 0,10) & 7 \\
 32 & 8710 & ( 2, 0,12) & ( 0, 2, 4) & ( 0, 0,32) & 3 \\
  \noalign{\smallskip}\hline
\end{tabular}
\hspace{2mm}
\begin{tabular}{r@{.}lc@{\hspace{0.3cm}}c@{\hspace{0.3cm}}cc}
 \hline\noalign{\smallskip} N & i & $\mathbf{L}^t_1$ &
 $\mathbf{L}^t_2$ & $\mathbf{L}^t_3$ & $\mathrm{I_G}$ \\
 \noalign{\smallskip}\hline\noalign{\smallskip}
 32 & 8709 & ( 2, 0,10) & ( 0, 2, 4) & ( 0, 0,32) & 5 \\
 32 & 8726 & ( 2, 0,10) & ( 0, 2, 6) & ( 0, 0,32) & 7 \\
 32 & 8745 & ( 2, 0,14) & ( 0, 2, 8) & ( 0, 0,32) & 7 \\
 32 & 10441 & ( 2, 0, 6) & ( 0, 4, 4) & ( 0, 0,16) & 7 \\
 32 & 10475 & ( 2, 0, 6) & ( 0, 4, 8) & ( 0, 0,16) & 7 \\
 32 & 8729 & ( 2, 0,16) & ( 0, 2, 6) & ( 0, 0,32) & 8 \\
 32 & 8744 & ( 2, 0,12) & ( 0, 2, 8) & ( 0, 0,32) & 6 \\
 32 & 10474 & ( 2, 0, 4) & ( 0, 4, 8) & ( 0, 0,16) & 6 \\
 32 & 10440 & ( 2, 0, 4) & ( 0, 4, 4) & ( 0, 0,16) & 8 \\
 32 & 8708 & ( 2, 0, 8) & ( 0, 2, 4) & ( 0, 0,32) & 10 \\
 32 & 14163 & ( 2, 2, 4) & ( 0, 8, 0) & ( 0, 0, 8) & 10 \\
 32 & 27780 & ( 4, 0, 4) & ( 0, 4, 4) & ( 0, 0, 8) & 10 \\
 32 & 14452 & ( 2, 4, 4) & ( 0, 8, 0) & ( 0, 0, 8) & 10 \\
 32 & 8711 & ( 2, 0,14) & ( 0, 2, 4) & ( 0, 0,32) & 10 \\
\noalign{\smallskip}\hline
\end{tabular}
\newpage

\section*{Appendix B. Dimensionless lowest energies,
$\epsilon_s$, per vertex and magnetization squared, $m^2_{x,s}$,
per vertex on finite lattice bcc lattices}

{\bf Table B1.} Lowest energies per vertex, $\epsilon_{1/2}$ and
$\epsilon_{3/2}$ for S = 1/2 and S = 3/2 and highest
magnetizations squared for S = 1/2 on odd-N bcc lattices

\vspace{2mm}
\begin{tabular}{r@{.}lccc}
\hline\noalign{\smallskip} N & i & $\epsilon_{1/2}$ &
$\epsilon_{3/2}$ & $m_{x,1/2}^2$ \\
\noalign{\smallskip}\hline\noalign{\smallskip}
 9 & 13  &  -1.111111  &  -1.000000 & 0.151235 \\
 11 & 15  &  -1.097845  &  -1.024081 & 0.145971 \\
 13 & 17  &  -1.089331  &  -1.036804 & 0.142017 \\
 13 & 18  &  -1.087852  &  -1.035348 & 0.142359 \\
 15 & 19  &  -1.084328  &  -1.045012 & 0.138698 \\
 15 & 20  &  -1.080184  &  -1.040916 & 0.139706 \\
 15 & 21  &  -1.081390  &  -1.042115 & 0.139452 \\
 15 & 28  &  -1.081370  &  -1.042097 & 0.139461 \\
 15 & 30  &  -1.082254  &  -1.042970 & 0.139267 \\
 15 & 38  &  -1.080958  &  -1.041686 & 0.139545 \\
 15 & 577  &  -1.080803  &  -1.041537 & 0.139598 \\
 17 & 22  &  -1.075262  &  -1.044788 & 0.137883 \\
 17 & 23  &  -1.074157  &  -1.043685 & 0.138107 \\
 17 & 24  &  -1.076656  &  -1.046173 & 0.137558 \\
 17 & 25  &  -1.076804  &  -1.046316 & 0.137506 \\
 19 & 24  &  -1.071814  &  -1.047477 & 0.135870 \\
 19 & 25  &  -1.069375  &  -1.045045 & 0.136405 \\
 19 & 26  &  -1.070551  &  -1.046218 & 0.136149 \\
 19 & 27  &  -1.073684  &  -1.049329 & 0.135368 \\
 19 & 37  &  -1.073588  &  -1.049235 & 0.135392 \\
 19 & 38  &  -1.071901  &  -1.047561 & 0.135846 \\
 21 & 26  &  -1.069336  &  -1.049447 & 0.134105 \\
 21 & 27  &  -1.066292  &  -1.046420 & 0.134862 \\
 21 & 28  &  -1.065706  &  -1.045831 & 0.134968 \\
 21 & 29  &  -1.070156  &  -1.050258 & 0.133858 \\
 21 & 30  &  -1.070145  &  -1.050248 & 0.133862 \\
 21 & 31  &  -1.071393  &  -1.051487 & 0.133500 \\
 21 & 38  &  -1.068786  &  -1.048905 & 0.134315 \\
 21 & 40  &  -1.068096  &  -1.048216 & 0.134435 \\
 21 & 41  &  -1.067288  &  -1.047416 & 0.134672 \\
 21 & 42  &  -1.066522  &  -1.046651 & 0.134830 \\
 21 & 49  &  -1.069085  &  -1.049199 & 0.134209 \\
 21 & 53  &  -1.071373  &  -1.051466 & 0.133509 \\
 21 & 73  &  -1.070043  &  -1.050148 & 0.133892 \\
 21 & 75  &  -1.066387  &  -1.046512 & 0.134834 \\
 21 & 86  &  -1.069112  &  -1.049225 & 0.134195 \\
 21 & 1332  &  -1.067424  &  -1.047550 & 0.134627 \\
\noalign{\smallskip}\hline
\end{tabular}
\hspace{1mm}
\begin{tabular}{r@{.}lccc}
\hline\noalign{\smallskip} N & i & $\epsilon_{1/2}$ &
$\epsilon_{3/2}$ & $m_{x,1/2}^2$ \\
\noalign{\smallskip}\hline\noalign{\smallskip}
 23 & 28  &  -1.067501  &  -1.050940 & 0.132537 \\
 23 & 29  &  -1.063762  &  -1.047226 & 0.133571 \\
 23 & 30  &  -1.063050  &  -1.046508 & 0.133671 \\
 23 & 31  &  -1.064184  &  -1.047640 & 0.133436 \\
 23 & 32  &  -1.069743  &  -1.053165 & 0.131821 \\
 23 & 33  &  -1.065691  &  -1.049142 & 0.133032 \\
 23 & 41  &  -1.066641  &  -1.050091 & 0.132890 \\
 23 & 45  &  -1.067099  &  -1.050543 & 0.132701 \\
 23 & 46  &  -1.063846  &  -1.047307 & 0.133546 \\
 23 & 53  &  -1.067190  &  -1.050632 & 0.132680 \\
 23 & 56  &  -1.069550  &  -1.052975 & 0.131895 \\
 25 & 30  &  -1.066090  &  -1.052086 & 0.131130 \\
 25 & 31  &  -1.061896  &  -1.047917 & 0.132392 \\
 25 & 32  &  -1.060664  &  -1.046687 & 0.132673 \\
 25 & 33  &  -1.061319  &  -1.047334 & 0.132442 \\
 25 & 34  &  -1.064092  &  -1.050099 & 0.131729 \\
 25 & 36  &  -1.062042  &  -1.048061 & 0.132336 \\
 25 & 44  &  -1.064898  &  -1.050907 & 0.131648 \\
 25 & 45  &  -1.064961  &  -1.050967 & 0.131639 \\
 25 & 49  &  -1.065471  &  -1.051473 & 0.131416 \\
 25 & 50  &  -1.062058  &  -1.048076 & 0.132346 \\
 25 & 58  &  -1.061775  &  -1.047798 & 0.132432 \\
 25 & 59  &  -1.065589  &  -1.051589 & 0.131353 \\
 25 & 72  &  -1.060732  &  -1.046755 & 0.132656 \\
 25 & 4227  &  -1.060729  &  -1.046751 & 0.132658 \\
 27 & 33  &  -1.060436  &  -1.048462 & 0.131329 \\
 27 & 34  &  -1.058698  &  -1.046732 & 0.131791 \\
 27 & 35  &  -1.059906  &  -1.047926 & 0.131353 \\
 27 & 36  &  -1.059721  &  -1.047747 & 0.131497 \\
 27 & 39  &  -1.059268  &  -1.047300 & 0.131647 \\
 27 & 49  &  -1.060493  &  -1.048519 & 0.131297 \\
 27 & 51  &  -1.060438  &  -1.048467 & 0.131391 \\
 27 & 52  &  -1.061408  &  -1.049434 & 0.131151 \\
 27 & 53  &  -1.060734  &  -1.048761 & 0.131285 \\
 27 & 54  &  -1.059340  &  -1.047370 & 0.131628 \\
 27 & 62  &  -1.060233  &  -1.048262 & 0.131408 \\
 27 & 66  &  -1.063687  &  -1.051699 & 0.130483 \\
 27 & 68  &  -1.059064  &  -1.047098 & 0.131711 \\
\noalign{\smallskip}\hline
\end{tabular}

\vspace{2mm}
\begin{tabular}{r@{.}lccc}
\hline\noalign{\smallskip} N & i & $\epsilon_{1/2}$ &
$\epsilon_{3/2}$ & $m_{x,1/2}^2$ \\
\noalign{\smallskip}\hline\noalign{\smallskip}
 27 & 81  &  -1.060187  &  -1.048216 & 0.131432 \\
 27 & 91  &  -1.063222  &  -1.051240 & 0.130677 \\
 27 & 94  &  -1.062741  &  -1.050755 & 0.130577 \\
 27 & 124  &  -1.059294  &  -1.047321 & 0.131584 \\
 27 & 136  &  -1.063248  &  -1.051266 & 0.130664 \\
 27 & 2354  &  -1.059580  &  -1.047612 & 0.131592 \\
 27 & 2396  &  -1.058818  &  -1.046855 & 0.131794 \\
 27 & 2399  &  -1.059844  &  -1.047868 & 0.131453 \\
 27 & 2549  &  -1.061395  &  -1.049421 & 0.131156 \\
 27 & 2550  &  -1.059735  &  -1.047765 & 0.131538 \\
 27 & 2594  &  -1.060740  &  -1.048769 & 0.131342 \\
 27 & 14309  &  -1.060109  &  -1.048142 & 0.131513 \\
 29 & 35  &  -1.059263  &  -1.048892 & 0.130361 \\
 29 & 36  &  -1.057294  &  -1.046932 & 0.130926 \\
 29 & 37  &  -1.057856  &  -1.047488 & 0.130734 \\
 29 & 38  &  -1.058752  &  -1.048376 & 0.130369 \\
 29 & 39  &  -1.059394  &  -1.049022 & 0.130310 \\
 29 & 41  &  -1.061719  &  -1.051336 & 0.129510 \\
 29 & 42  &  -1.057354  &  -1.046991 & 0.130912 \\
 29 & 50  &  -1.062484  &  -1.052101 & 0.129494 \\
 29 & 51  &  -1.062027  &  -1.051648 & 0.129725 \\
 29 & 52  &  -1.058575  &  -1.048210 & 0.130630 \\
 29 & 55  &  -1.062289  &  -1.051908 & 0.129609 \\
\noalign{\smallskip}\hline
\end{tabular}
\hspace{2mm}
\begin{tabular}{r@{.}lccc}
\hline\noalign{\smallskip} N & i & $\epsilon_{1/2}$ &
$\epsilon_{3/2}$ & $m_{x,1/2}^2$ \\
\noalign{\smallskip}\hline\noalign{\smallskip}
 29 & 57  &  -1.058853  &  -1.048486 & 0.130533 \\
 29 & 58  &  -1.057423  &  -1.047059 & 0.130894 \\
 29 & 66  &  -1.058950  &  -1.048582 & 0.130483 \\
 29 & 73  &  -1.059037  &  -1.048667 & 0.130458 \\
 29 & 85  &  -1.062309  &  -1.051928 & 0.129597 \\
 31 & 37  &  -1.058318  &  -1.049248 & 0.129470 \\
 31 & 38  &  -1.056109  &  -1.047048 & 0.130164 \\
 31 & 39  &  -1.056223  &  -1.047161 & 0.130125 \\
 31 & 40  &  -1.057808  &  -1.048734 & 0.129468 \\
 31 & 41  &  -1.056782  &  -1.047715 & 0.129914 \\
 31 & 44  &  -1.058351  &  -1.049280 & 0.129450 \\
 31 & 45  &  -1.056228  &  -1.047166 & 0.130120 \\
 31 & 54  &  -1.060975  &  -1.051898 & 0.128894 \\
 31 & 55  &  -1.057395  &  -1.048331 & 0.129846 \\
 31 & 58  &  -1.057387  &  -1.048323 & 0.129871 \\
 31 & 59  &  -1.061060  &  -1.051982 & 0.128849 \\
 31 & 61  &  -1.057905  &  -1.048838 & 0.129643 \\
 31 & 62  &  -1.056273  &  -1.047210 & 0.130115 \\
 31 & 71  &  -1.061290  &  -1.052210 & 0.128746 \\
 31 & 74  &  -1.057821  &  -1.048754 & 0.129694 \\
 31 & 78  &  -1.057971  &  -1.048902 & 0.129631 \\
 31 & 94  &  -1.055874  &  -1.046815 & 0.130243 \\
\noalign{\smallskip}\hline
\end{tabular}

\vspace{8mm}
\noindent {\bf Table B2.} Lowest energies per vertex, $\epsilon_0$
and $\epsilon_1$ for S = 0 and S = 1 and highest magnetizations
squared for S = 0 on even-N bcc lattices

\begin{tabular}{r@{.}lccc}
\hline\noalign{\smallskip} N & i & $\epsilon_0$ & $\epsilon_1$ &
$m_{x,0}^2$ \\ \noalign{\smallskip}\hline\noalign{\smallskip}
 10 & 15 & -1.115022  &  -1.070219 & 0.149610 \\
 12 & 25 & -1.099498  &  -1.068599 & 0.145017  \\
 12 & 17 & -1.100778  &  -1.069868 & 0.144744  \\
 14 & 21 & -1.090423  &  -1.067827 & 0.141360  \\
 14 & 20 & -1.089310  &  -1.066715 & 0.141581  \\
 14 & 19 & -1.092238  &  -1.069629 & 0.140898  \\
 16 & 23 & -1.081668  &  -1.064433 & 0.139034  \\
 16 & 1642 & -1.078975  &  -1.061714 & 0.139531  \\
 16 & 41 & -1.083024  &  -1.065785 & 0.138724  \\
 16 & 24 & -1.082957  &  -1.065720 & 0.138750  \\
 16 & 32 & -1.083102  &  -1.065864 & 0.138698  \\
 16 & 21 & -1.086794  &  -1.069535 & 0.137714  \\
 18 & 37 & -1.076639  &  -1.063062 & 0.136877  \\
 18 & 25 & -1.074930  &  -1.061351 & 0.137208  \\
 18 & 1101 & -1.076270  &  -1.062694 & 0.136951  \\
 18 & 2024 & -1.073869  &  -1.060277 & 0.137399  \\
 18 & 24 & -1.076782  &  -1.063203 & 0.136829  \\
 18 & 26 & -1.077395  &  -1.063814 & 0.136668  \\
 18 & 35 & -1.077087  &  -1.063509 & 0.136777  \\
 18 & 36 & -1.078914  &  -1.065327 & 0.136301  \\
 18 & 1001 & -1.077833  &  -1.064251 & 0.136573  \\
 18 & 27 & -1.077572  &  -1.063989 & 0.136612  \\
 18 & 34 & -1.078467  &  -1.064883 & 0.136401  \\
 18 & 67 & -1.078797  &  -1.065210 & 0.136273  \\
 \noalign{\smallskip}\hline
\end{tabular}
\hspace{2mm}
\begin{tabular}{r@{.}lccc}
\hline\noalign{\smallskip} N & i & $\epsilon_0$ & $\epsilon_1$ &
$m_{x,0}^2$ \\ \noalign{\smallskip}\hline\noalign{\smallskip}
 20 & 27 & -1.070541  &  -1.059569 & 0.135571  \\
 20 & 50 & -1.071442  &  -1.060473 & 0.135405  \\
 20 & 30 & -1.071890  &  -1.060915 & 0.135254  \\
 20 & 62 & -1.069670  &  -1.058694 & 0.135723  \\
 20 & 630 & -1.072857  &  -1.061883 & 0.135087  \\
 20 & 2446 & -1.069862  &  -1.058882 & 0.135678  \\
 20 & 26 & -1.073220  &  -1.062244 & 0.134948  \\
 20 & 38 & -1.073013  &  -1.062039 & 0.135028  \\
 20 & 61 & -1.075150  &  -1.064166 & 0.134404  \\
 20 & 2447 & -1.069906  &  -1.058925 & 0.135662  \\
 20 & 73 & -1.073071  &  -1.062096 & 0.135005  \\
 20 & 629 & -1.072875  &  -1.061902 & 0.135078  \\
 20 & 29 & -1.075189  &  -1.064205 & 0.134393  \\
 20 & 40 & -1.075014  &  -1.064032 & 0.134456  \\
 20 & 63 & -1.075106  &  -1.064122 & 0.134428  \\
 20 & 51 & -1.075166  &  -1.064181 & 0.134409  \\
 22 & 29 & -1.067179  &  -1.058131 & 0.134195  \\
 22 & 33 & -1.067353  &  -1.058303 & 0.134137  \\
 22 & 31 & -1.069031  &  -1.059977 & 0.133699  \\
 22 & 43 & -1.068563  &  -1.059512 & 0.133873  \\
 22 & 30 & -1.066438  &  -1.057386 & 0.134304  \\
 22 & 2908 & -1.066695  &  -1.057639 & 0.134236  \\
 22 & 41 & -1.069958  &  -1.060905 & 0.133547  \\
 22 & 28 & -1.070624  &  -1.061566 & 0.133285  \\
 22 & 66 & -1.070377  &  -1.061321 & 0.133389  \\
 \noalign{\smallskip}\hline
\end{tabular}

\vspace{2mm}
\begin{tabular}{r@{.}lccc}
\hline\noalign{\smallskip} N & i & $\epsilon_0$ & $\epsilon_1$ &
$m_{x,0}^2$ \\ \noalign{\smallskip}\hline\noalign{\smallskip}
 22 & 69 & -1.072611  &  -1.063546 & 0.132684  \\
 22 & 42 & -1.070401  &  -1.061344 & 0.133389  \\
 22 & 32 & -1.072775  &  -1.063709 & 0.132625  \\
 22 & 44 & -1.069958  &  -1.060905 & 0.133547  \\
 24 & 30 & -1.068651  &  -1.061048 & 0.131805  \\
 24 & 31 & -1.064665  &  -1.057074 & 0.132955  \\
 24 & 32 & -1.063751  &  -1.056158 & 0.133125  \\
 24 & 33 & -1.064305  &  -1.056710 & 0.132992  \\
 24 & 34 & -1.069432  &  -1.061826 & 0.131549  \\
 24 & 36 & -1.063865  &  -1.056275 & 0.133138  \\
 24 & 44 & -1.067580  &  -1.059983 & 0.132255  \\
 24 & 45 & -1.068202  &  -1.060602 & 0.131999  \\
 24 & 46 & -1.065818  &  -1.058222 & 0.132573  \\
 24 & 47 & -1.064946  &  -1.057355 & 0.132915  \\
 24 & 48 & -1.065731  &  -1.058140 & 0.132746  \\
 24 & 49 & -1.065218  &  -1.057626 & 0.132816  \\
 24 & 57 & -1.068174  &  -1.060573 & 0.132021  \\
 24 & 58 & -1.064527  &  -1.056937 & 0.133002  \\
 24 & 59 & -1.068478  &  -1.060878 & 0.131977  \\
 24 & 61 & -1.070772  &  -1.063162 & 0.131121  \\
 24 & 71 & -1.067526  &  -1.059929 & 0.132284  \\
 24 & 73 & -1.065990  &  -1.058392 & 0.132524  \\
 24 & 86 & -1.063825  &  -1.056235 & 0.133149  \\
 24 & 87 & -1.064988  &  -1.057394 & 0.132854  \\
 24 & 88 & -1.066667  &  -1.059070 & 0.132372  \\
 24 & 100 & -1.063459  &  -1.055869 & 0.133222  \\
 24 & 113 & -1.067629  &  -1.060032 & 0.132236  \\
 24 & 873 & -1.067678  &  -1.060081 & 0.132218  \\
 24 & 875 & -1.068331  &  -1.060732 & 0.132032  \\
 24 & 888 & -1.066408  &  -1.058814 & 0.132557  \\
 24 & 894 & -1.067585  &  -1.059988 & 0.132259  \\
 24 & 1860 & -1.065748  &  -1.058156 & 0.132743  \\
 24 & 1911 & -1.068268  &  -1.060669 & 0.132057  \\
 24 & 2705 & -1.066061  &  -1.058468 & 0.132619  \\
 24 & 3410 & -1.064190  &  -1.056594 & 0.132986  \\
 24 & 3411 & -1.064043  &  -1.056447 & 0.133043  \\
 24 & 3424 & -1.063994  &  -1.056399 & 0.133061  \\
 24 & 4266 & -1.063990  &  -1.056395 & 0.133064  \\
 24 & 3412 & -1.064285  &  -1.056688 & 0.132950  \\
 24 & 3425 & -1.064697  &  -1.057097 & 0.132787  \\
 26 & 33 & -1.062746  &  -1.056285 & 0.131835 \\
 26 & 34 & -1.061196  &  -1.054738 & 0.132228 \\
 26 & 35 & -1.062356  &  -1.055891 & 0.131810 \\
 \noalign{\smallskip}\hline
\end{tabular}
\hspace{2mm}
\begin{tabular}{r@{.}lccc}
\hline\noalign{\smallskip} N & i & $\epsilon_0$ & $\epsilon_1$ &
$m_{x,0}^2$ \\ \noalign{\smallskip}\hline\noalign{\smallskip}
 26 & 36 & -1.063033  &  -1.056570 & 0.131737 \\
 26 & 38 & -1.064957  &  -1.058490 & 0.131140 \\
 26 & 39 & -1.061341  &  -1.054882 & 0.132183 \\
 26 & 47 & -1.065798  &  -1.059332 & 0.131066 \\
 26 & 48 & -1.065874  &  -1.059407 & 0.131037 \\
 26 & 49 & -1.062641  &  -1.056180 & 0.131875 \\
 26 & 50 & -1.062636  &  -1.056176 & 0.131878 \\
 26 & 51 & -1.065896  &  -1.059429 & 0.131026 \\
 26 & 52 & -1.065720  &  -1.059254 & 0.131108 \\
 26 & 61 & -1.066418  &  -1.059948 & 0.130806 \\
 26 & 76 & -1.065989  &  -1.059521 & 0.131004 \\
 26 & 78 & -1.064163  &  -1.057699 & 0.131481 \\
 26 & 92 & -1.063033  &  -1.056570 & 0.131737 \\
 26 & 122 & -1.061352  &  -1.054894 & 0.132167 \\
 26 & 5912 & -1.062218  &  -1.055753 & 0.131869  \\
 26 & 5913 & -1.061799  &  -1.055336 & 0.132039  \\
 28 & 6785 & -1.059855  &  -1.054290 & 0.131187  \\
 28 & 6786 & -1.060332  &  -1.054765 & 0.130993  \\
 28 & 6816 & -1.060055  &  -1.054489 & 0.131109  \\
 28 & 6784 & -1.060669  &  -1.055100 & 0.130850  \\
 28 & 6800 & -1.059807  &  -1.054242 & 0.131206  \\
 30 & 7717 & -1.058308  &  -1.053465 & 0.130397  \\
 30 & 7718 & -1.058525  &  -1.053681 & 0.130307  \\
 30 & 7733 & -1.058113  &  -1.053270 & 0.130473  \\
 30 & 7734 & -1.058524  &  -1.053680 & 0.130309  \\
 30 & 7750 & -1.058751  &  -1.053906 & 0.130211  \\
 30 & 7716 & -1.059447  &  -1.054599 & 0.129908  \\
 30 & 10754 & -1.058525  &  -1.053681 & 0.130309  \\
 32 & 8710 & -1.056918  &  -1.052665 & 0.129722  \\
 32 & 8709 & -1.057082  &  -1.052828 & 0.129654  \\
 32 & 8726 & -1.056915  &  -1.052663 & 0.129500  \\
 32 & 8745 & -1.057128  &  -1.052874 & 0.129604  \\
 32 & 10441 & -1.056711  &  -1.052459 & 0.129724  \\
 32 & 10475 & -1.057036  &  -1.052783 & 0.129636  \\
 32 & 8729 & -1.057217  &  -1.052963 & 0.129805  \\
 32 & 8744 & -1.057455  &  -1.053200 & 0.129670  \\
 32 & 10474 & -1.057218  &  -1.052964 & 0.129604  \\
 32 & 10440 & -1.057957  &  -1.053701 & 0.129270  \\
 32 & 8708 & -1.058465  &  -1.054207 & 0.129035  \\
 32 & 14163 & -1.057220  &  -1.052966 & 0.129602  \\
 32 & 27780 & -1.056805  &  -1.052552 & 0.129438  \\
 32 & 14452 & -1.057632  &  -1.053376 & 0.129770  \\
 32 & 8711 & -1.057914  &  -1.053658 & 0.128926  \\
 \noalign{\smallskip}\hline
\end{tabular}

\newpage

\end{document}